\documentclass[useAMS,usenatbib]{mn2e}
\pdfoutput=1 
\usepackage{amsmath}
\usepackage{graphicx}
\usepackage{txfonts}
\usepackage{natbib}

\newcommand{\teff}{T_{\rm{eff}}}
\newcommand{\teffsun}{T_{\rm{eff},\odot}}
\newcommand{\logg}{\log\,g}
\newcommand{\feh}{\rm{[Fe/H]}}
\newcommand{\aFe}{[\alpha/\rm{Fe}]}
\newcommand{\ergs}{\rm{erg\,s^{-1}\,cm^{-2}\,\AA^{-1}}}
\newcommand{\ergHz}{\rm{erg\,s^{-1}\,cm^{-2}\,Hz^{-1}}}
\newcommand{\BB}{\rm{Bol}}
\newcommand{\kms}{\rm{km\,s^{-1}}}
\newcommand{\fbol}{f_{\rm Bol}}
\newcommand\msol{{\cal M_{\odot}}}

\newcommand\lta{\mathrel{\hbox{\raise 0.6 ex \hbox{$<$}\kern
                   -1.8 ex\lower .5 ex\hbox{$\sim$}}}}
\newcommand\gta{\mathrel{\hbox{\raise 0.6 ex \hbox{$>$}\kern
                   -1.7 ex\lower .5 ex\hbox{$\sim$}}}}

\title[Synthetic photometry for stellar models]{Synthetic Stellar Photometry -- I. General considerations and new transformations for broad-band systems.}

\author[Casagrande \& VandenBerg]{\parbox{18cm}{
    L.~Casagrande$^{1}$\thanks{Stromlo Fellow}\thanks{Email:luca.casagrande@anu.edu.au}, 
  Don A.~VandenBerg$^{2}$}\\
  $^1$ Research School of Astronomy and Astrophysics, Mount Stromlo 
  Observatory, The Australian National University, ACT 2611, Australia\\
  $^2$ Department of Physics \& Astronomy, University of Victoria, P.O.~Box 
  1700 STN CSC, Victoria, BC, V8W 2Y2, Canada}

\begin{document}

\date{Received; accepted}

\maketitle

\begin{abstract}
After a pedagogical introduction to the main concepts of synthetic photometry, 
colours and bolometric corrections in the Johnson-Cousins, 2MASS, and 
HST-ACS/WFC3 photometric systems are generated from MARCS synthetic fluxes
for various $\feh$ and $\aFe$ combinations, and virtually any value of $E(B-V)
\lesssim 0.7$. 
The successes and failures of model fluxes in reproducing the 
observed magnitudes are highlighted. Overall, extant synthetic fluxes predict 
quite realistic broad-band colours and bolometric corrections, especially at 
optical and longer wavelengths: further improvements of the predictions for 
the blue and ultraviolet spectral regions await the use of hydrodynamic models 
where the microturbulent velocity is not treated as a free parameter. 
We show how the morphology of the colour-magnitude diagram (CMD) changes for 
different values of $\feh$ and $\aFe$; in particular,how suitable colour 
combinations can easily discriminate between red giant branch and lower main 
sequence populations with different $\aFe$, due to the concomitant loops and 
swings in the CMD. We also provide computer programs to produce tables of 
synthetic bolometric corrections as well as routines to interpolate in them. 
These colour--$\teff$--metallicity relations may be used to convert isochrones 
for different chemical compositions to various bandpasses assuming observed 
reddening values, thus bypassing the standard assumption of a constant colour 
excess for stars of different spectral type. We also show how such an 
assumption can lead to significant systematic errors. The MARCS transformations 
presented in this study promise to provide important constraints on our
understanding of the multiple stellar populations found in globular clusters
(e.g., the colours of lower main sequence stars are predicted to depend 
strongly on $\aFe$) and of those located towards/in the Galactic bulge.
\end{abstract}


\section{Introduction}

Photometric systems and filters are designed to be sensitive temperature,
gravity, or metal abundance indicators and thereby to complement spectroscopic
determinations of the fundamental properties of stars.  Moreover, 
when studying more complex systems such as star clusters and galaxies, the
integrated magnitudes and colours of stars can be used to infer the ages,
metallicities, and other properties of the underlying stellar populations.
To accomplish this, filter systems are tailored to select regions in stellar 
spectra where the variations of the atmospheric parameters leave their 
characteristic traces with enough prominence to be detected in photometric
data.  Beginning with the influential papers by \cite{j66} and \cite{s66},
which describe the basis of broad- and intermediate-band photometry, a large
number of systems exists nowadays, and more are being developed with the
advent of extensive photometric surveys \citep[see e.g.,][for a
review]{bessell05}.

Broad-band colours are usually tightly correlated with the stellar 
effective temperature ($\teff$), although metallicity ($\feh$) and (to a lesser
extent) surface gravity ($\logg$) also play a role, especially towards the 
near ultraviolet (UV) and the Balmer discontinuity 
\citep[e.g.,][]{els62,ridgway80,bell89,alonso96:teff_scale,ivezic08,c10}.  On 
the other hand, intermediate- and/or narrow-band filters centred on specific 
spectral features can have a much higher sensitivity to stellar parameters 
other than $\teff$ \citep[e.g.,][]{s66,wing67,mc68,golay72,ms82,w94}.  While 
broad-band photometry can be easily used to map and study extended stellar 
populations and/or large fractions of the sky, also at faint magnitudes
\citep[e.g.,][]{shs98,bedin04,ivezic07,saito12}, intermediate- and narrow-band
photometry is more limited in this respect, although still very informative 
\citep[e.g.,][]{mb82,bb82,yong08,arnadottir10,c14a}.

In principle, determining stellar parameters from photometric data is a basic 
task, yet empirical calibrations connecting them are generally limited to
certain spectral types and/or involve substantial observational work. 
In recent years, we have made a considerable effort to derive empirical 
relations that link photometric indices to the effective temperatures and 
metallicities of dwarf and subgiant stars \citep{c10,c11}.  In particular, our
studies of solar twins have enabled us to accurately set the zero-points of 
both 
the $\feh$ and $\teff$ scales for the first time \citep{melendez10,r12,c12}  
\citep[Similar work is currently underway for giants, see also][]{c14b}.
The zero-point of
photometrically derived $\teff$s is also intimately connected to the absolute
calibration of photometric systems; i.e., it is crucial to convert magnitudes
into fluxes and vice versa, be they synthetic or observed. 

Parallel to this empirical approach, in the spirit of \cite{vc03} and 
\cite{cvg04}, we have also tested the performance of synthetic colours against 
field and cluster stars of different evolutionary stages \citep{vcs10,bsv10}. 
Such comparisons, which have been carried out using the Victoria-Regina
models \citep{vbd12} and an initial set of colour transformations based on
one-dimensional MARCS synthetic stellar fluxes \citep{g08}, have validated the 
accuracy of the adopted library in predicting broad-band colours for a wide
range of $\teff$ and $\feh$ relevant to dwarf stars and subgiants.  Some
discrepancies between the predicted and observed colours of giants were found,
but they could be telling us that the model temperatures are not realistic
because of deficiencies in, e.g., the treatment of convection or the 
atmospheric boundary condition.  It is also possible that these differences 
stem from abundance anomalies in some of the globular clusters that were used
in these studies.

\begin{figure}
\begin{center}
\includegraphics[width=0.45\textwidth]{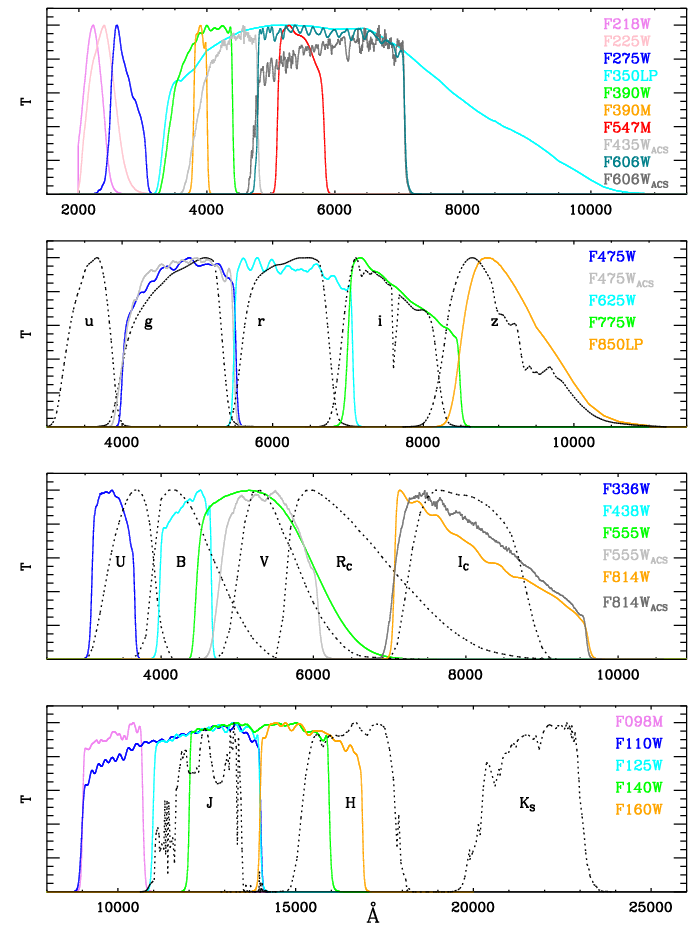}
\caption{System response functions from which synthetic 
colours have been computed. All curves are normalized to one, and 
shown in arbitrary units as function of wavelength (in \AA). 
For the HST system, different colours correspond to the names on 
the right hand side of each panel (WFC3, unless the subscript indicates 
the ACS camera). Sloan (second), Johnson-Cousins (third) and 2MASS (fourth 
panel) system response functions are plotted with dotted lines and the name of 
each filter is indicated next to its curve. Some of the HST filters are 
designed to broadly match the Sloan and Johnson-Cousins systems: compare dotted 
and continuous lines in the second and third panel.}\label{f:filters}
\end{center}
\end{figure}

In this paper, we continue our efforts to provide reliable MARCS synthetic 
colours for different $\feh$ and $\aFe$ combinations in the most commonly used 
optical and infrared (IR) photometric systems (Figure \ref{f:filters}).  
In particular, we take full 
advantage of the accuracy we have established concerning the absolute 
calibration of photometric systems to compute synthetic colours with well 
defined zero-points. 
Also, for the first time, we generate synthetic colours accounting for
different values of the reddening, $E(B-V)$, thus bypassing the usual
assumption that the colour excess is independent of spectral type.  Our
purpose here is thus twofold: together with assessing the performance of the 
latest MARCS grids of synthetic fluxes in different bands, we provide
several computer programs (in Fortran) to produce, and interpolate in, tables
of bolometric corrections (BCs, see Appendix \ref{app}).  The resultant 
colour--$\teff$--metallicity relations are given for the most commonly used
broad-band systems.

\section{Synthetic photometry}\label{sec:sp}

The basic idea of synthetic photometry is to reproduce observed colours 
(be they stars, planets, or galaxies) based simply on input (usually 
theoretical) spectra $f_x$, and a characterization of the instrumental 
system response function $T_\zeta$ under which the photometric observations 
are performed (i.e., the total throughput of the optics, detector, 
filter, \dots over a spectral range $x_A$ to $x_B$. If observations are 
ground-based, then the atmospheric transmission must also be included in the 
throughput: this is often done at a nominal airmass of $1.3$). Thus, in its 
most general 
and simplest form, a synthetic magnitude can be written as proportional 
(via a logarithmic factor, see Section \ref{sec:fm}) to 
$\int_{x_A}^{x_B} f_x\,T_\zeta\,{\rm d}x$, where the integration over $x$ 
is carried out either in wavelength or frequency space. 
This relation also
highlights the fact that a magnitude can be simply thought of as an
heterochromatic measurement, where the information encoded in the spectrum is
weighted by $T_\zeta$\footnote{Because synthetic magnitudes are essentially
a weighted average of the flux, the correct formalism requires a normalization
over $T_\zeta$, thus 
$\frac{\int_{x_A}^{x_B} f_x\,T_\zeta\,{\rm d}x}{\int_{x_A}^{x_B}
T_\zeta\,{\rm d}x}$.}.  Depending on the spectral coverage and the
transmissivity of $T_\zeta$, the result will depend more or less strongly upon
(some of) the physical parameters of the underlying spectrum.  By employing
different filter combinations, it is then possible to highlight certain
spectral features.   If we are interested in stellar colours, these could be, 
for instance, the slope of the continuum, the depression owing to metal lines,
the Balmer discontinuity, etc., which in turn tell us about the basic stellar 
parameters.

The advantage of using synthetic libraries is that they cover uniformly and 
homogeneously a range of stellar parameters --- a feature which is appealing
for a number of reasons including, in particular, the transposition of 
theoretical stellar models from the $\teff$-luminosity plane onto the 
observational colour-magnitude diagram (CMD).  Several papers deal with 
this topic, and excellent discussions on synthetic photometry can be found in 
e.g., \cite{b98}, \cite{g02} and \cite{bm12}.

\subsection{From fluxes to magnitudes}\label{sec:fm}

Even though the physical quantities that we work with are fluxes, in
astronomy it is customary to deal with magnitudes, a concept that is thought
to have originated with Hipparchus.  In his system, the brightest stars in the
sky, first magnitude, were considered to be about twice as bright as those of
second magnitude, and so forth until the faintest stars visible with the naked
eye, which are about one hundred times fainter, were classified as sixth
magnitude.  This concept was formalized by \cite{pogson}, who conveniently
proposed that a first magnitude star be $100^{1/5} \simeq 2.5$ times as bright
as a second magnitude star.  This underpins the {\it definition} of magnitudes
as proportional to $-2.5 \log$ of the flux, i.e.,
\begin{equation}\label{eq:pogson}
m_1 - m_2 = -2.5 \log \frac{\int_{x_A}^{x_B} f_{x,1}\,T_\zeta\,{\rm d}x}{\int_{x_A}^{x_B} f_{x,2}\,T_\zeta\,{\rm d}x}
\end{equation}
where in the case of Pogson $T_\zeta$ was the transmission of the eye, and 
using $m_2$ and $f_{x,2}$ as standard star (or the Hipparchus ``first magnitude 
stars''), the heterochromatic definition of a photometric system is readily 
obtained (see e.g., Eq. \ref{eq:vega}). The fact 
that a non-linear transformation relates magnitudes to fluxes, is a potential 
source of bias, which we deal with in Appendix \ref{bias}. Here 
it suffices to mention that this bias arises when using poor quality data, and 
assuming that Gaussian random errors in fluxes are mapped into Gaussian random 
errors in magnitudes, and vice versa.

\subsection{Photon counting versus Energy integration}

Before dealing with specific magnitude systems one further concept must 
be introduced.
The system response function $T_\zeta$ represents the total throughput in
reaching the observer, which is affected by everything lying between a 
photon as it arrives at the top of the Earth's atmosphere and its final
detection \citep[e.g.,][]{golay74}. A proper characterization of $T_\zeta$ is
therefore non-trivial, although nowadays it is accurately known for most 
photometric systems.  Among the different stages and components which define
the total throughput, the most important ones are arguably the filter itself
and the response function of the detector.  The latter must be taken into
account in the response function of the overall system because, in all
instances, the flux is evaluated after entering the detector.  Because of
this, one must distinguish between photon-counting (e.g., CCD) and
energy-integration (e.g., photo-multiplier) detectors 
\citep[e.g.,][]{bessell2000,apellaniz06,bm12}. 

For a CCD system, the rate of arrival of photons (number per unit time per
detector area) from a source having a flux $f_\lambda$ is 
$\int \frac{f_\lambda}{h\nu}\,T_\zeta\,\rm{d}\lambda$, while the energy
measured in the same window of time over the same detector area is simply 
$\int f_\lambda\,T_\zeta\,\rm{d}\lambda$. Since a CCD counts photons (or 
actually converts photons into electrons with a given efficiency), it follows 
that $\int \frac{f_\lambda}{h\nu}\,T_\zeta\,{\textrm d}\lambda = \frac{1}{hc}
\int f_\lambda\,\lambda\,T_\zeta\,\rm{d}\lambda\,$; i.e., the energy
formalism can also be used for a CCD, if $T_\zeta$ is replaced by
$\lambda\,T_\zeta$ (or equivalently, $T_\zeta$ by $\frac{T_\zeta}{\nu}$ if
working in frequency space).  A common source of confusion when using
published response functions for different photometric systems is whether or
not $T_\zeta$ has already been multiplied by $\lambda$.  Throughout this paper,
the formalism that we introduce for heterochromatic measurements always assumes
that $T_\zeta$ is provided in energy integration form and that all synthetic
colours are generated for photon-counting systems; i.e., we explicitly introduce
the $\lambda\,T_\zeta$ (or $\frac{T_\zeta}{\nu}$) term in the equations of 
synthetic photometry. (Note that the $hc$ constant disappears because of the 
normalization over $T_\zeta$). Since energy integration response functions 
are the correct ones to use in the photon-counting formalism, they are 
sometimes also called photonic response functions \citep[e.g.,][]{bm12}.
Among the existing photometric packages, the most commonly used one is 
Synphot (Laidler et al.~2008)\nocite{synphot}, which comply with the 
photon-counting formalism (and photonic response functions) described
here\footnote{http://www.stsci.edu/institute/software\_hardware/stsdas/synphot
and its python version http://stsdas.stsci.edu/pysynphot}. Needless to say,
in a number of cases where the published values of $T_\zeta$ already include
the $\lambda$ dependence (see Table \ref{t:pq}), the photon-counting synthetic
quantities have been calculated using the energy-integration formalism.

\subsection{{\tt ST} mag system}\label{sec:ST}

{\tt ST} monochromatic magnitudes (i.e.~per unit wavelength) are defined  as
\begin{equation}
m_{\tt ST}=-2.5 \log f_{\lambda} + ZP_{\lambda}
\end{equation}
where, by construction, a constant flux density per unit wavelength 
$f_{\lambda}^{0}=3.631 \times 10^{-9}\,\ergs$ is defined to have
$m_{\tt ST}=0.0$, thus implying $ZP_{\lambda}=-21.10$.  Since in reality all
measurements are heterochromatic, it follows that {\tt ST} magnitudes are in
fact defined over the wavelength range $\lambda_i$ to $\lambda_{f}$ which
characterizes a given bandpass $\zeta$ that has a system response function
$T_{\zeta}$
\begin{equation}\label{eq:st_het}
m_{\tt ST}=-2.5 \log \frac{\int_{\lambda_i}^{\lambda_f} \lambda f_{\lambda}
 T_{\zeta} 
\rm{d}\lambda} {\int_{\lambda_i}^{\lambda_f} \lambda
 T_{\zeta} \rm{d}\lambda}-21.10 = 
-2.5 \log \frac{\int_{\lambda_i}^{\lambda_f} \lambda f_{\lambda}
 T_{\zeta} \rm{d}\lambda} {f_{\lambda}^{0} \int_{\lambda_i}^{\lambda_f}
 \lambda T_{\zeta} \rm{d}\lambda}.
\end{equation}
To express the above $\lambda$-formalism in terms of a constant flux density 
per unit frequency, the only change that needs to be made to
Eq.~(\ref{eq:st_het}) is to replace $f_{\lambda}^{0}$ by
$f_{\nu}^{0}\frac{c}{\lambda^2}$.

\subsection{{\tt AB} mag system}
Conceptually identical to the {\tt ST} magnitudes, the {\tt AB} system is 
defined as
\begin{equation}\label{eq:ab_mono}
m_{\tt AB}=-2.5 \log f_{\nu} + ZP_{\nu}
\end{equation}
where, in this case, $m_{\tt AB}=0.0$ corresponds to a constant flux density
per unit frequency $f_{\nu}^{0}=3.631\,\times 10^{-20} \ergHz$, thus implying
$ZP_{\nu}=-48.60$.  Integrating over frequencies (heterochromatic measurement)
\begin{equation}\label{eq:ab_het}
m_{\tt AB}=-2.5 \log \frac{\int_{\nu_i}^{\nu_f}f_{\nu}\frac{T_{\zeta}}{\nu}\rm{d}\nu}
{\int_{\nu_i}^{\nu_f}\frac{T_\zeta}{\nu}\rm{d}\nu}-48.60 =
-2.5 \log \frac{\int_{\nu_i}^{\nu_f}f_{\nu}
 T_\zeta \rm{d}\ln\nu}{f_{\nu}^{0}\int_{\nu_i}^{\nu_f}T_\zeta \rm{d}\ln\nu}
\end{equation}
which is identical to, e.g., the photon-counting definition given by
\citet{fuku96}.  Recasting Eq.~(\ref{eq:ab_het}) in terms of wavelength
results in
\begin{displaymath}
m_{\tt AB}=-2.5 \log \frac{\int_{\lambda_i}^{\lambda_f} \lambda f_{\lambda}
 T_\zeta \rm{d}\lambda}
{f_{\nu}^{0} c \int_{\lambda_i}^{\lambda_f}
 \frac{T_\zeta}{\lambda} \rm{d}\lambda} =
\end{displaymath}
\begin{equation}
-2.5 \log \frac{\int_{\lambda_i}^{\lambda_f} \lambda f_{\lambda}
 T_\zeta \rm{d}\lambda}
{f_{\lambda}^{0} \int_{\lambda_i}^{\lambda_f} \lambda
 T_\zeta \rm{d}\lambda} - 2.5 \log 
\frac{f_{\lambda}^{0}}{f_{\nu}^{0} c} -
 2.5 \log \frac{\int_{\lambda_i}^{\lambda_f} \lambda T_\zeta
 \rm{d}\lambda}{\int_{\lambda_i}^{\lambda_f}\frac{T_\zeta}{\lambda}\rm{d}\lambda}
\end{equation}
where the dimensionless quantity 
$-2.5 \log \frac{f_{\lambda}^{0}}{f_{\nu}^{0} c} = 18.6921$.  With
$\lambda_{PIVOT,\zeta} = \left(\frac{\int \lambda T_\zeta \rm{d} \lambda}{\int
 \frac{T_\zeta}{\lambda} \rm{d}\lambda}\right)^{\frac{1}{2}}$,
we obtain
\begin{equation}\label{eq:abst}
m_{\tt AB} = m_{\tt ST} - 5 \log \lambda_{PIVOT,\zeta} + 18.6921.
\end{equation}
From Eq.~(\ref{eq:abst}) it follows that {\tt AB} and {\tt ST} magnitudes are 
identical at a wavelength $\simeq 5475$~\AA.  As \cite{og83} chose to use 
the absolute flux of $\alpha$\,Lyr (Vega) at $5480$~\AA~to define the
wavelength\footnote{Note that \cite{og83} adopted $V=0.03$ 
for $\alpha$\,Lyr, from which $m_{\tt AB} \sim V$ once the absolute flux at 
$5556$~\AA~measured by \cite{hl75} $3.50\,\times 10^{-20} \ergHz$ was adopted. 
For an object with a relatively flat spectrum, this implied that $m_{\tt AB}=0$
at $f_{\nu}^{0}$ i.e. close to the value of $3.65\,\times 10^{-20} \ergHz$ 
measured by \cite{os70} at $5480$~\AA~for a star of $V=0.0$.} at which 
$m_{\tt AB}=m_{\tt VEGA}$, the flux values $f_{\nu}^{0}$ and $f_{\lambda}^{0}$
are chosen so that, for convenience, $\alpha$\,Lyr has very similar magnitudes
in all systems, {\tt ST}, {\tt AB}, and {\tt VEGA}.  Note that a revision of
the actual flux scale of $\alpha$\,Lyr (or any network of spectrophotometric
standards) affects only the way that photometric observations are standardized
to those systems, not their definitions.  (That is, when standardizing
observations, zero-point shifts can be applied in order to exactly match the
original definitions).

\subsection{{\tt VEGA} mag system}\label{sec:VEGA}

This system, which is the most well known one, uses $\alpha$\,Lyr (Vega) as the
primary calibrating star, as in the case of the most famous and still widely
used Johnson-Cousins system (and many others).  Historically, the zero-points
of the Johnson system were defined ``in terms of unreddened main-sequence (MS)
stars of class A0 \dots with an accuracy sufficient to permit the placement
of the zero-point to about $0.01$ mag'' \citep{jm53}. While observationally
this zero-point is often defined by a network of standard stars, formally it
can be anchored to just one object, the usual choice being $\alpha$\,Lyr. 
The definition of the {\tt VEGA} system is given only for heterochromatic 
measurements 
\begin{equation}\label{eq:vega}
m_{\tt VEGA} = -2.5 \log \frac{\int_{\lambda_i}^{\lambda_f}
 \lambda f_{\lambda} T_\zeta \rm{d}\lambda}{\int_{\lambda_i}^{\lambda_f}
 \lambda T_\zeta \rm{d}\lambda} + ZP_\zeta
\end{equation}
where $ZP_\zeta$ is derived for each bandpass $\zeta$ using a star of known 
absolute flux $f_{\lambda}=f_\star$ and observed magnitude 
$m_{\tt VEGA}=m_{\star,\zeta}$ 
\citep[e.g.,][where our adopted magnitudes for Vega are reported in the 
third column of Table \ref{t:pq}]{b98,g02,c06}. Thus 
\begin{equation}
ZP_\zeta = m_{\star,\zeta} + 2.5 \log \frac{\int_{\lambda_i}^{\lambda_f}
 \lambda f_{\star} 
T_\zeta \rm{d}\lambda}{\int_{\lambda_i}^{\lambda_f}
 \lambda T_\zeta \rm{d}\lambda} = 
m_{\star,\zeta} + 2.5 \log \bar{f}_{\star,\zeta}
\end{equation}
where $\bar{f}_{\star,\zeta}$ is the absolute calibration of the photometric 
system in the $\zeta$ band.  $ZP_\zeta$ is thus of crucial importance for 
the synthetic photometry that we want to generate.  To achieve the correct
standardization, it is possible to modify either the adopted absolute
calibration or the standard magnitude observed in the $\zeta$ band (or both).
When generating synthetic magnitudes, an evaluation of both quantities must be
provided (see Table \ref{t:pq}).

As already noted, $\alpha$\,Lyr is commonly used as the zero-point standard. 
Should its theoretical spectrum $F_{\lambda}$ (see next Section) be modelled 
with sufficient precision, and its angular diameter $\theta$ accurately
measured, its absolute flux would be straightforward to obtain. In practice,
its pole-on and rapidly rotating nature, as well as its suspected variability 
\citep[although longward of $12\micron$,][and thus of no impact on the colours 
investigated here]{aumann84} complicate things \cite[also see the discussion
in][]{c06}. In practice, the best approach resorts to the use of a composite
absolutely calibrated spectrum for different wavelength regions. Arguably,
the best absolute flux $f_{\star}$ for $\alpha$\,Lyr is currently available
from the CALSPEC library\footnote{Regularly updated absolute spectrophotometry 
can be found at http://www.stsci.edu/hst/observatory/cdbs/calspec.html. 
If not otherwise specified, we have used the STIS005 spectrum in this 
investigation.}, which intermingles 
{\it Hubble Space Telescope} (HST) absolute spectrophotometry in the range 
$1675 - 5350$ \AA, with an especially tailored model flux longward of this 
wavelength. The overall accuracy of this absolute flux is $\sim 1$\% 
\citep{bohlin07}, which translates to about $0.01$~mag in $ZP_\zeta$. If not 
otherwise specified, this is the absolute flux that we have used when
generating synthetic magnitudes in the {\tt VEGA} system.

\subsection{Reality Check}

It follows from the previous sections that the formalism to define a 
photometric system is sound, independently of whether the {\tt ST, AB,} or 
{\tt VEGA} system is used. However, its actual realization at the telescope 
is far from being simple \citep[see also][]{bessell05}. Factoring in all 
possible sources of uncertainty in terms of the quantities we are discussing 
here means that the absolute flux, as well as the magnitudes, of the 
standard star might not be accurate (nor precise) enough (as is probably the
case for $\alpha$\,Lyr), and/or the actual system transmission curves might not
be in exact agreement with those tabulated.  In fact, more often than not,
(small) zero-point corrections are still needed to replicate the definition of
a given photometric system as closely as possible (and more generally, 
linear terms rather than zero-points might even be more appropriate, even 
though they are rarely, if ever, considered). That is, for certain
filters, a correction term $\epsilon_\zeta$ might be added to the right-hand 
sides of Eq.~(\ref{eq:st_het}), (\ref{eq:ab_het}) and (\ref{eq:vega}).
Table \ref{t:pq} reports when such corrections are necessary. It is worth 
mentioning that these corrections depend on the definitions adopted for each 
photometric system; consequently, different authors might use opposite signs. 
The values listed here comply with the definitions adopted in the present 
paper.
\begin{table*}
\centering
\caption{Characteristic parameters defining the photometric systems studied 
here. The absolute calibration $\bar{f}_{\star,\zeta}$ (obtained using a
spectrum of Vega as defined in Section \ref{sec:VEGA})
is given in $\ergs$.}\label{t:pq}
\begin{tabular}{c|l|ccc|c|c}
\hline 
        &            &       \multicolumn{3}{|c|}{{\tt VEGA} system}    & {\tt AB} system     & {\tt ST} system     \\
Filter  & $T_\zeta$   &            &                 &             &               &               \\
        &            & $m_{\star,\zeta}$ & $\bar{f}_{\star,\zeta}$      & $\epsilon_\zeta$  & $\epsilon_\zeta$    & $\epsilon_\zeta$    \\
\hline
$UX$       &   $1^\ast$ & $+0.020$     & 4.0275E-9       &     $0$     &     $-$       &    $-$ \\
$BX$       &   $1^\ast$ & $+0.020$     & 6.2712E-9       &     $0$     &     $-$       &    $-$ \\
 $B$       &   $1^\ast$ & $+0.020$     & 6.3170E-9       &     $0$     &     $-$       &    $-$ \\
 $V$       &   $1^\ast$ & $+0.030$     & 3.6186E-9       &     $0$     &     $-$       &    $-$ \\
 $R_C$     &   $1^\ast$ & $+0.039$     & 2.1652E-9       &     $0$     &     $-$       &    $-$ \\
 $I_C$     &   $1^\ast$ & $+0.035$     & 1.1327E-9       &     $0$     &     $-$       &    $-$ \\
 $U$       &   2       & $-$         & $-$             &     $-$     &     $-0.761$  &    $-$ \\
 $B$       &   2       & $-$         & $-$             &     $-$     &     $+0.130$  &    $-$ \\
 $V$       &   2       & $-$         & $-$             &     $-$     &     $+0.020$  &    $-$ \\
 $R_C$     &   2       & $-$         & $-$             &     $-$     &     $-0.163$  &    $-$ \\
 $I_C$     &   2       & $-$         & $-$             &     $-$     &     $-0.404$  &    $-$ \\
 $J$       &   $3^\ast$ & $-0.001$    & 3.129E-10       &  $-0.017$   &     $-$       &    $-$ \\
 $H$       &   $3^\ast$ & $+0.019$    & 1.133E-10       &  $+0.016$   &     $-$       &    $-$ \\
 $K_S$     &   $3^\ast$ & $-0.017$    & 4.283e-11       &  $+0.003$   &     $-$       &    $-$ \\
 $u$               &   4       & $-$         & $-$             &     $-$     &     $+0.037$  &    $-$ \\
 $g$               &   4       & $-$         & $-$             &     $-$     &     $-0.010$  &    $-$ \\
 $r$               &   4       & $-$         & $-$             &     $-$     &     $+0.003$  &    $-$ \\
 $i$               &   4       & $-$         & $-$             &     $-$     &     $-0.006$  &    $-$ \\
 $z$               &   4       & $-$         & $-$             &     $-$     &     $-0.016$  &    $-$ \\
 $F435W_{\rm{ACS}}$ &   5       & $0$         & 6.4530E-9       &     $0$     &     $0$  &    $0$ \\
 $F475W_{\rm{ACS}}$ &   5       & $0$         & 5.3128E-9       &     $0$     &     $0$  &    $0$ \\
 $F555W_{\rm{ACS}}$ &   5       & $0$         & 3.8191E-9       &     $0$     &     $0$  &    $0$ \\
 $F606W_{\rm{ACS}}$ &   5       & $0$         & 2.8705E-9       &     $0$     &     $0$  &    $0$ \\
 $F814W_{\rm{ACS}}$ &   5       & $0$         & 1.1353E-9       &     $0$     &     $0$  &    $0$ \\
 $F218W$           &   6      & $0$         & 4.6170E-9       &     $0$     &     $0$  &    $0$ \\
 $F225W$           &   6      & $0$         & 4.1887E-9       &     $0$     &     $0$  &    $0$ \\
 $F275W$           &   6      & $0$         & 3.7273E-9       &     $0$     &     $0$  &    $0$ \\
 $F336W$           &   6      & $0$         & 3.2449E-9       &     $0$     &     $0$  &    $0$ \\
 $F350LP$          &   6      & $0$         & 2.7464E-9       &     $0$     &     $0$  &    $0$ \\
 $F390M$           &   6      & $0$         & 6.5403E-9       &     $0$     &     $0$  &    $0$ \\
 $F390W$           &   6      & $0$         & 5.7932E-9       &     $0$     &     $0$  &    $0$ \\
 $F438W$           &   6      & $0$         & 6.6896E-9       &     $0$     &     $0$  &    $0$ \\
 $F475W$           &   6      & $0$         & 5.2240E-9       &     $0$     &     $0$  &    $0$ \\
 $F547M$           &   6      & $0$         & 3.6694E-9       &     $0$     &     $0$  &    $0$ \\
 $F555W$           &   6      & $0$         & 3.9511E-9       &     $0$     &     $0$  &    $0$ \\
 $F606W$           &   6      & $0$         & 2.9094E-9       &     $0$     &     $0$  &    $0$ \\
 $F625W$           &   6      & $0$         & 2.4383E-9       &     $0$     &     $0$  &    $0$ \\
 $F775W$           &   6      & $0$         & 1.3104E-9       &     $0$     &     $0$  &    $0$ \\
 $F814W$           &   6      & $0$         & 1.1486E-9       &     $0$     &     $0$  &    $0$ \\
 $F850LP$          &   6      & $0$         & 8.0146E-10      &     $0$     &     $0$  &    $0$ \\
 $F098M$           &   6      & $0$         & 6.6687E-10      &     $0$     &     $0$  &    $0$ \\
 $F110W$           &   6      & $0$         & 4.0648E-10      &     $0$     &     $0$  &    $0$ \\
 $F125W$           &   6      & $0$         & 3.0448E-10      &     $0$     &     $0$  &    $0$ \\
 $F140W$           &   6      & $0$         & 2.0837E-10      &     $0$     &     $0$  &    $0$ \\
 $F160W$           &   6      & $0$         & 1.4553E-10      &     $0$     &     $0$  &    $0$ \\
\hline
\end{tabular}
\begin{minipage}{1\textwidth}
An asterisk in the second column indicates that $T_\zeta$ given in the reference 
is already multiplied by $\lambda$ and renormalized.
--1: Bessell (1990).\,
--2: Bessell \& Murphy (2012): notice that the formalism introduced in that 
paper for the $UBV(RI)_C$ system is in terms of {\tt AB} magnitudes, although 
the system is actually {\tt VEGA} based; $\epsilon_\zeta$ values are obtained 
from their tables 3 and 5 (but with the opposite sign, following our 
definition of $\epsilon_\zeta$).\,
--3: Cohen et al.~(2003); $\epsilon_\zeta$ values from Casagrande et 
al.~(2010).\,
--4: SDSS website; $\epsilon_\zeta$ values are obtained comparing the observed 
magnitudes of a number of HST CALSPEC standards with the magnitudes generated
in the {\tt AB} system using the adopted $T_\zeta$ with the HST absolute 
spectrophotometry.\,
--5: http://www.stsci.edu/hst/acs/analysis/zeropoints.\, 
--6: http://www-int.stsci.edu/$\sim$WFC3/UVIS/SystemThroughput.
\end{minipage}
\end{table*}

\subsection{Bolometric corrections}

Libraries of synthetic stellar spectra provide the flux on a surface element 
$F_{\lambda}$ (usually in $\ergs$) for a given set of stellar parameters (see 
Section \ref{s:lib}) from which it is straightforward to compute the
theoretical absolute flux 
\begin{equation}\label{eq:fl}
f_\lambda=10^{-0.4 \rm{A}_\lambda} \left( \frac{R}{d} \right)^2 F_\lambda, 
\end{equation}
for a star of radius $R$, at a distance $d$, and (if present) suffering from 
interstellar extinction of magnitude $\rm{A}_\lambda$ at wavelength $\lambda$. 
Such theoretical determinations of $f_\lambda$ can then be inserted into the
formalism outlined in the previous sections in order to compute synthetic
magnitudes.  Often only synthetic colours (i.e., the difference in the
magnitudes measured through two filters) are computed, since doing so cancels 
out the dependence on the dilution factor $\frac{R}{d} = \frac{\theta}{2}$. 
Similarly, if one wishes to compute the BC in a given band 
$BC_\zeta$
\begin{equation}\label{eq:bc0}
BC_\zeta = m_{\BB} - m_\zeta = M_{\BB} - M_\zeta
\end{equation}
where the lower and upper cases refer to apparent and absolute magnitudes, 
respectively. For simplicity, when dealing with synthetic fluxes, we use the 
same dilution factor $R_{\odot}/d_{10}$ for all models, where 
$d_{10}=10 \rm{pc}$, i.e.\,for the solar spectrum we also compute its absolute 
magnitudes. For any models it follows that
\begin{equation}
M_\zeta = -2.5 \log \left[ \left(\frac{R_{\odot}}{d_{10}}\right)^2 \mathcal{K}_\zeta 
\right] + zp_\zeta
\end{equation}
where the exact form of $zp_\zeta$ (which now includes $\epsilon_\zeta$, if 
necessary) and $\mathcal{K}_\zeta$ (i.e., the integration of $f_\lambda$ over 
the system response function) depends whether synthetic magnitudes are computed
according to Eq.~(\ref{eq:st_het}), (\ref{eq:ab_het}) or (\ref{eq:vega}). 
Also, from the definition of the absolute bolometric magnitude 
\begin{equation}\label{eq:bcdef}
M_{\BB} = -2.5 \log \frac{R^2\teff^4}{R_{\odot}^2\teffsun^2} + M_{\BB,\odot},
\end{equation}
because of the above choice on the dilution factor, the BC then becomes
\begin{equation}\label{eq:bc}
BC_\zeta = -2.5 \log \left( \frac{\teff}{\teffsun} \right)^4 + M_{\BB,\odot}
 + 2.5 \log \left[ \left(\frac{R_\odot}{d_{10}}\right)^2 \mathcal{K} \right]
 - zp_\zeta.
\end{equation}

In this paper we have adopted $M_{\BB,\odot}=4.75$, but it should be 
appreciated that this value is simply a {\it definition}, not a measurement.
In fact, BCs were originally defined in the visual only, 
in such a way as to be negative for all stars \citep{kui38}. From the spectral 
energy distribution, we expect BCs in the visual to be 
largest for very hot and for very cool stars. The minimum would then 
occur around spectral type F, which then would set the zero-point of the 
bolometric magnitudes. For the Sun, this implied a value $BC_{V,\odot}$ between 
$-0.11$ \citep{aller63} and $-0.07$ \citep{ma68}. However, with 
the publication of a larger grid of model atmospheres, e.g.,~the smallest 
BC in the Kurucz's grid (1979) \nocite{kurucz79} 
implied $BC_{V,\odot} = -0.194$. These difficulties can instead be avoided by 
adopting a fixed zero-point, which is currently the preferred choice. 
Therefore, any value of $M_{\BB,\odot}$ is equally legitimate, on the condition 
that once chosen, all BCs are rescaled accordingly \citep[e.g.,][]{torres10}. 
As a matter of fact, whenever bolometric magnitudes are published, be they 
apparent or absolute, the adopted $M_{\BB,\odot}$ should always be 
specified\footnote{It follows from Eq.~(\ref{eq:bcdef}) that adopting 
$M_{\BB,\odot}=4.75$ with the GONG collaboration solar luminosity value 
$3.846 \times 10^{26}$~W, a star with $M_{\BB}=0$ has a radiative luminosity of 
$3.055 \times 10^{28}$~W, which is the flux zero-point of the BC scale 
according to the definitions adopted by IAU Commissions 25 and 36 at the 
IAU 1997 General Assembly.}.
We also note
that using the solar values reported in Table \ref{t:sun}, our choice implies
$-0.07 \lesssim BC_{V,\odot} \lesssim -0.06$.

\subsection{Caveats with extinction}\label{sec:ext}

Before dealing with extinction, it is useful to clarify the wavelength with 
which heterochromatic measurements (magnitudes in this case) are associated.
Any heterochromatic measurement can always be reduced to a monochromatic flux
$\bar{f}$ (also called effective flux) via
\begin{equation}
\bar{f} = \frac{\int_{\lambda_i}^{\lambda_f} \lambda f_{\lambda}
 T_\zeta \rm{d}\lambda}{\int_{\lambda_i}^{\lambda_f} \lambda
 T_\zeta \rm{d}\lambda}. 
\end{equation}
Assuming that the function $f_{\lambda}$ is continuous and that $T_\zeta$ does 
not change sign in the interval ($\lambda_i, \lambda_f$), the generalization of 
the mean value theorem states that there is at least one value of $\lambda$ in 
the above interval such that
\begin{equation}
f_{\tilde{\lambda}} \int_{\lambda_i}^{\lambda_f} \lambda T_\zeta \rm{d}\lambda = \int_{\lambda_i}^{\lambda_f} \lambda f_{\lambda} T_\zeta \rm{d}\lambda.
\end{equation}
Rearranging this, we obtain
\begin{equation}
\bar{f}=f_{\tilde{\lambda}} = \frac{\int_{\lambda_i}^{\lambda_f} \lambda f_{\lambda} T_\zeta \rm{d}\lambda}{\int_{\lambda_i}^{\lambda_f} \lambda T_\zeta \rm{d}\lambda}
\end{equation}
where $\tilde{\lambda}$ is the isophotal wavelength. The latter is thus the 
wavelength which must be attached to the monochromatic quantity $\bar{f}$
that is obtained from a heterochromatic measurement.  Stellar spectra do not
necessarily satisfy the requirement of the mean value theorem for integration,
as they exhibit discontinuities.  Although the mean value of the intrinsic
flux is well defined, the determination of the isophotal wavelength becomes
problematic because spectra contain absorption lines, and hence the definition
of $\tilde{\lambda}$ can yield multiple solutions \citep[e.g.,][]{tv05,rieke08}.
It is possible to avoid these complications by introducing the effective
wavelength (here in photon-counting formalism)
\begin{equation}
\lambda_{eff} = \frac{\int_{\lambda_i}^{\lambda_f} \lambda^2 f_{\lambda}
 T_\zeta \rm{d}\lambda}{\int_{\lambda_i}^{\lambda_f} \lambda f_{\lambda}
 T_\zeta \rm{d}\lambda}
\end{equation}
i.e., the mean wavelength of the passband weighted by the energy distribution 
of the source over that band, which is a good approximation for 
$\tilde{\lambda}$ \citep{golay74}. The effective wavelength thus shifts with 
$f_\lambda$, depending on the source under investigation since it depends on
both the extinction and the intrinsic stellar spectrum (Eq.~\ref{eq:fl}). If
present, extinction is usually parameterized as follows 
$A_\lambda = R_V\,E(B-V) \left[ a(\lambda^{-1}) + b(\lambda^{-1})/R_V\right]$, 
where $E(B-V)$ is the colour excess (or reddening) and 
$R_V \equiv \frac{A_V}{E(B-V)}$ is the ratio of total to selective extinction 
in the optical, assumed to be constant $\simeq 3.1$ for most line of sights 
\citep[e.g.,][]{ccm89}. The above parametrization is known as the extinction 
law. Note that the extinction law is derived primarily using early O- and 
B-type stars, i.e. it is expressed in terms of an $A_V = R_V\,E(B-V)$
relation that is valid for these spectral types.
Also, despite being given in a monochromatic form (i.e.~with a $\lambda$ 
dependence), it is actually derived using photometric (i.e.~heterochromatic) 
quantities, which are valid at the effective wavelengths appropriate for the 
stars plus filters used to derive the extinction law. While the implications of 
these facts are 
discussed in the literature \citep[e.g.,][]{ftz99,McCall04,ma13}, their 
importance is often overlooked. We call 
the $R_V$ and $E(B-V)$ values which enter the extinction law ``nominal''.
The amount of attenuation in a given bandpass $\zeta$ (i.e.~the magnitude 
difference, where the subscript $0$ indicates an unreddened source) is then
\begin{equation}
m_\zeta-m_{\zeta,0} = -2.5 \log \frac{\int_{\lambda_i}^{\lambda_f}
 \lambda 10^{-0.4 A_\lambda} F_\lambda
 T_\zeta \rm{d}\lambda}{\int_{\lambda_i}^{\lambda_f} \lambda F_\lambda
 T_\zeta \rm{d}\lambda} = A_\zeta
\end{equation}
and a given nominal $E(B-V)$ and $R_V$ will produce a different effective flux 
associated with a different effective wavelength, depending on the intrinsic 
spectral energy distribution $F_\lambda$ of the source under investigation. 
In other words, the colour excess between two bands 
$E(\zeta-\eta) = A_\zeta - A_\eta$ 
will vary with spectral class as well as with $E(B-V)$ and $R_V$. 
For example, assuming $R_V=3.1$, the ratio $E(u-z)/E(B-V)$ in a metal-poor 
subgiant star ($\feh=-3.0$, $\logg=3.5, \teff=6500$~K) will vary from 
$3.333$ to $3.319$ assuming a nominal $E(B-V)=0.01$ or $0.70$, respectively. 
The same ratio will be $3.283$ or $3.275$ in a solar metallicity giant 
($\logg=2.0, \teff=4000$~K), and $3.375$ or $3.362$ in a solar metallicity M 
dwarf ($\logg=5.0, \teff=2500$~K).  If $R_V=2.5$, which seems to be
representative of selected directions towards the bulge \citep{nataf12}, the 
aforementioned numbers would change to $3.188$ and $3.164$ (metal-poor
subgiant), $3.123$ and $3.108$ (solar metallicity giant), and $3.211$ and
$3.187$ (solar metallicity M dwarf).
\begin{figure*}
\begin{center}
\includegraphics[width=0.9\textwidth]{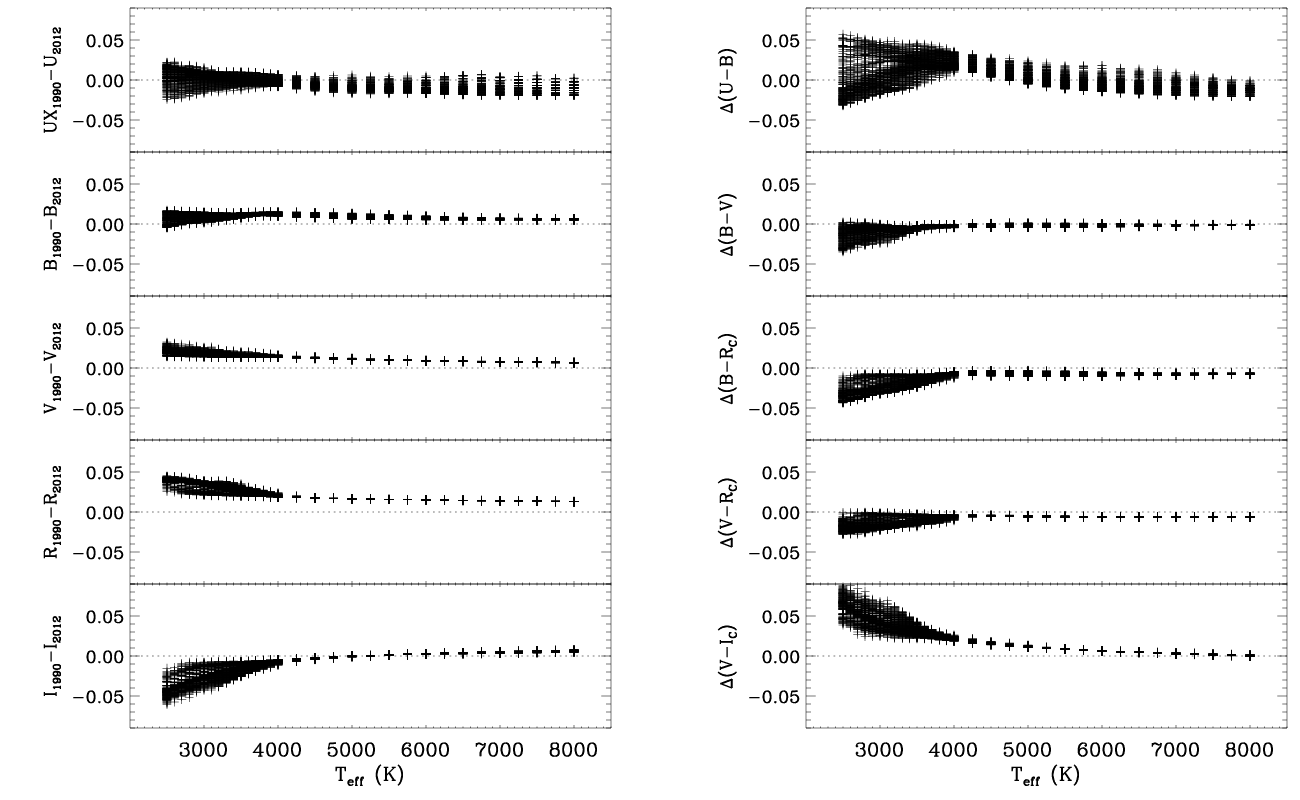}
\caption{{\it Left panels:} difference between the {\tt ubvri90} and 
{\tt ubvri12} magnitudes. Each point is a model flux of different 
$\teff,\,\logg$ and $\feh$, according to the values available from the MARCS 
library ($\alpha$-standard composition). {\it Right panels:} same difference 
for selected colour indices.}
\label{f:b90b12}
\end{center}
\end{figure*}

Based on the above considerations, it is also important to stress 
that the nominal $R_V$ and $E(B-V)$ values entering the extinction law --- as 
derived from early type stars --- change when moving towards later 
spectral types as a result of the change in effective wavelength.
As an example, in the case of a late type star, the effective wavelengths of
the $B$ and $V$ filters shift to longer wavelengths, where the extinction
decreases, lowering both $A_B$ and $A_V$ for a given amount of dust.  As a
result, $R_V=\frac{A_V}{E(B-V)}=\frac{A_V}{A_B-A_V}$ increases (even if the
amount and physical properties of the dust are the same!) because the decrease
in the denominator of this ratio overwhelms the decrease of the numerator.
Thus, it is important to take this effect into account to 
distinguish whether a variation in $R_V$ stems from shifts in the effective 
wavelengths or from changes in the nature of the dust \citep[see e.g., the 
excellent discussion in][]{McCall04}.
For reference, a nominal $R_V=3.1$ and $E(B-V)=0.60$ in the aforementioned 
extinction law return $R_V=3.4$ and $E(B-V) = 0.55$ for a moderately 
metal-poor turn-off star. (We will elaborate on our discussion of this
practical example in Section \ref{sec:morpho} and Appendix \ref{app}.)

These subtleties are usually neglected when, e.g., isochrones are fitted to the
CMDs of halo globular clusters that have low values of the reddening, but they 
might not be negligible if the object under consideration is in a heavily
obscured region of the Galactic disc or bulge.  Furthermore, when using an
$E(B-V)$ value from the literature, it should be kept in mind how this value
was obtained; i.e., whether it is based on stars \citep[and if so, which
spectral types, e.g,][]{sf11}, or even background galaxies 
\citep[e.g.,][]{sfd98}.  Indeed, because of the above issues, it should not
come as a surprise to find that small adjustments to literature values of
$E(B-V)$ are needed to obtain a satisfactory fit an isochrone to a given CMD
(though this is just one of many possible causes of discrepancies between
predicted and observed colours). In addition, as we already pointed out,
the use of a constant colour excess across the entire CMD is not optimal in 
the presence of high extinction.  We discuss in Section \ref{sec:morpho} how
the set of reddened colours provided in this investigation enables one to
address these subtleties. 

\subsection{Johnson-Cousins $UBV(RI)_C$ and 2MASS $JHK_S$}\label{subsec:jc2m} 

Synthetic colours in the Johnson-Cousins and 2MASS filter sets have been 
computed in the {\tt VEGA} system only, and they have already been used in
the investigations by \cite{vcs10} and \cite{bsv10}, where further details
can be found.  Briefly, the absolute calibration of the Johnson-Cousins system
is obtained using the updated CALSPEC absolute spectrophotometry\footnote{At 
the time of generating those magnitudes, the most updated spectrum 
available was the STIS003, which underpins the absolute calibration of the 
Johnson-Cousins system reported in Table \ref{t:pq}. Should e.g.,~the STIS005 
spectrum be used, the Johnson-Cousins absolute fluxes in Table \ref{t:pq} 
would 
change into $UX$\,=\,4.0027E-9, $BX$\,=\,6.2313E-9, $B$\,=\,6.2763E-9, and
$V$\,=\,3.6112E-9 $\ergs$, thus implying magnitude differences of $0.007$ in 
$UX, BX$, and $B$ and $0.002$ in $V$ band. The fluxes in $R_C$ and $I_C$ would 
remain identical.} 
of $\alpha$\,Lyr \citep{bohlin07} and the filter transmission curves of
\cite{b90a}: the results are forced to match the observed magnitudes of Vega
\citep{b90b}. Because of problems standardizing the $U$ magnitudes, \cite{b90a}
provides two transmission curves for the $B$ system: one dubbed ``$BX$'' to be
used in conjunction with ``$UX$'' for the computation of the $U-B$ colour
index, and the other labeled ``$B$'' for use with any of the redder bandpasses 
$V(RI)_C$ to calculate, e.g., $B-V$ or $V-I_C$ colours.  We have chosen to use
the name {\tt ubvri90} to identify these magnitudes.

An additional set of synthetic Johnson-Cousins colours has also been computed 
following the prescriptions of \cite{bm12}, i.e., using improved filter 
transmission curves for the $U$ and $B$ bands (thereby removing the need to 
distinguish between, e.g., ``BX'' and ``B'') and newly derived zero-points.
The novel approach of \cite{bm12} is that the Johnson-Cousins system is
formulated in terms of {\tt AB} magnitudes, despite being on the {\tt VEGA}
system. We have named these 
magnitudes {\tt ubvri12}.  It is very comforting to find that the differences 
between the two sets of Johnson-Cousins magnitudes are relatively minor and
essentially constant, never exceeding $0.02$~mag ($0.05$~mag) above 
(below) $4000$~K. The same is also true when colour indices are compared
instead of magnitudes; with regard to $BVR_C$, in particular, the constant
offsets found in magnitudes conspire to give nearly the same colours above
$4000$~K (see Figure \ref{f:b90b12}).

For the 2MASS $JHK_S$ system, the $\alpha$~Lyr magnitudes, absolute 
calibration, and system response functions of \cite{cohen03} are used, 
including the small zero-point corrections found in \cite{c10}. The 
difference with respect to the CALSPEC absolute flux in the near-infrared 
is minor \cite[see discussion in][]{c10}, and our choice is motivated by the 
fact that the 2MASS absolute calibration adopted here has been thoroughly 
tested \citep[][]{rieke08,c12}.  Note that both \cite{b90a} and \cite{cohen03}
system response functions are photon-counting, i.e., they are already
multiplied by $\lambda$ and renormalized.  Thus, $T_\zeta$ (not
$\lambda\,T_\zeta$) must be used in the equations of synthetic photometry
(Section \ref{sec:sp}). With these choices, the absolute calibration in these
filters is essentially identical to that underlying the $\teff$ scale of
\cite{c10} and verified via analyses of solar twins\footnote{The only
difference is the $BV(RI)_C$ absolute calibration in \cite{c10}, which is
based on earlier HST spectrophotometry of $\alpha$\,Lyr \citep[as reported in 
Table A2 of][]{c06}, producing $\teff$ values that differ by only a few 
Kelvin.  However, this difference essentially cancels out when the solar twins 
are used in the calibrating process; i.e., the absolute calibration derived 
from \cite{bohlin07} agrees with that based on the solar twins.}. 

\subsection{Sloan Digital Sky Survey $ugriz$}

SDSS photometry is defined to be on the {\tt AB} system, although small 
post-facto zero-point corrections are necessary \citep{eisen06,hb06}. We have 
generated {\tt AB} magnitudes using the SDSS filter transmission 
curves\footnote{http://www.sdss3.org/instruments/camera.php\#Filters} and 
determined the necessary zero-point corrections $\epsilon_\zeta$ using stars 
with absolute flux measurements from the CALSPEC library and having SDSS 
$ugriz$ measurements (Casagrande et al., in prep.). 

Two remarks must be made: first, the exact definition of SDSS 
magnitudes is on the asinh system, where the classical Pogson logarithmic 
scale (Section \ref{sec:fm}) is replaced by the inverse hyperbolic sine 
function \citep[``asinh magnitudes'', see][]{lupton99}. Conveniently, this 
definition is virtually 
identical to the standard Pogson magnitudes at high signal--to--noise, but it 
behaves better at magnitudes fainter than the detection repeatability limit 
(i.e.~much fainter than $20^{\rm{th}}$~mag for SDSS). Brighter than this 
threshold the difference between the asinh and Pogson magnitudes is negligible 
and rapidly goes to zero. This makes the Pogson formulation outlined so far 
fully appropriate for generating synthetic magnitudes, besides being more
convenient given the dependence of the asinh 
magnitudes\footnote{https://www.sdss3.org/dr10/algorithms/magnitudes.php\#asinh}
on a softening parameter (survey dependent) and the difficulty of defining 
quantities such as bolometric corrections and absolute magnitudes in the asinh 
formulation \citep[see e.g.,][]{ggoc}.

The second remark concerns the potential confusion between the {\it primed} 
$u'g'r'i'z'$ and {\it unprimed} $ugriz$ magnitudes. The basis of the 
photometric calibration of the SDSS is defined by the 158 primary standards 
measured by the USNO 40-inches telescope \citep{smith02}, which however has 
filters with different effective wavelengths than those of the SDSS $2.5\rm{m}$ 
survey telescope\footnote{see http://www.sdss.org/dr7/algorithms/fluxcal.html
for an extensive explanation}. The latter defines the {\it unprimed} $ugriz$ 
system for which photometry has been published from SDSS DR1 onward, and for 
which we provide synthetic colours in this work. {\it Primed} magnitudes 
should thus be transformed into the {\it unprimed} system before comparing 
and/or using them with the synthetic photometry published here \citep[see 
e.g.,][]{tkr06}.

\subsection{HST}

The distinctive feature of {\it Hubble Space Telescope} (HST) photometry is
the use of the {\tt AB} and {\tt ST} systems, in addition to the {\tt VEGA}
system. We have generated BCs and colours for a number of
filters commonly used in two instruments: the Advanced Camera for Surveys
(ACS) is a third generation HST instrument which has provided some of the 
deepest photometric optical images ever obtained \citep{sbc07}, while the 
Wide Field Camera 3 (WFC3) is a fourth generation instrument (replacing the
Wide Field Planetary Camera 2, WFPC2) that covers the electromagnetic spectrum 
from UV to the IR.  For both cameras, a large number of filters is available,
covering long-pass ($LP$), broad ($W$), intermediate ($M$) and narrow ($N$)
bands.  Some of them closely match such commonly used filters as those in the
Johnson-Cousins, SDSS, Str\"omgren, and Washington systems 
\citep[e.g.,][]{bessell05}. To provide
continuity with previous observations, several of the WFC3 filters are very
similar to those used in the WFPC2 and ACS instruments
\citep[e.g.,][]{wfc3handbook}. We have produced transformations in the 
{\tt AB}, {\tt ST} and {\tt VEGA} systems for the following filters: $F435W$, 
$F475W$, $F555W$, $F606W$ and $F814W$ (ACS camera); $F218W$, $F225W$, $F275W$,
$F336W$, $F350LP$, $F390M$, $F390W$, $F438W$, $F475W$, $F547M$, $F555W$,
$F606W$, $F625W$, $F775W$, $F814W$, $F850LP$ (WFC3 optical) and $F098M$,
$F110W$, $F125W$, $F140W$, $F160W$ (WFC3 infrared). 

Updated values for the system response functions as well as regular 
improvements to the absolute flux of $\alpha$\,Lyr are provided by the 
HST team.  We list in Table \ref{t:pq} our adopted values. The absolute
calibration of $\alpha$\,Lyr that we derived for the ACS system differs
slightly from the values currently listed in the 
website\footnote{http://www.stsci.edu/hst/acs/analysis/zeropoints/\#vega}.
These differences in absolute flux correspond to zero-point shifts of 
$\simeq0.01$~mag in $F435W$ and $<0.005$~mag in the other ACS filters 
considered here. At any given time, users can readily compute the difference 
in magnitude implied by the $\bar{f}_\star$ that we adopted, and they may
use any updated/preferred value, shifting if necessary our computed colours
by this amount.

We conclude this section with a cautionary note: particular care should be
taken when working with stars that have cool effective temperatures and/or
high value of reddening (the exact values of these quantities depend on the
filter being used). Some of the UV filters are known to have red leaks towards
longer wavelengths.  Although these leaks are usually confined to much less 
than 1\%, they may not be negligible in some cases \cite[see table 6.5
in][]{wfc3handbook}, which could cause the synthetic colours to be anomalous 
and the effective wavelengths to be considerably redder than the values
expected from the nominal filter bandpasses.
\begin{figure*}
\begin{center}
\includegraphics[width=0.8\textwidth]{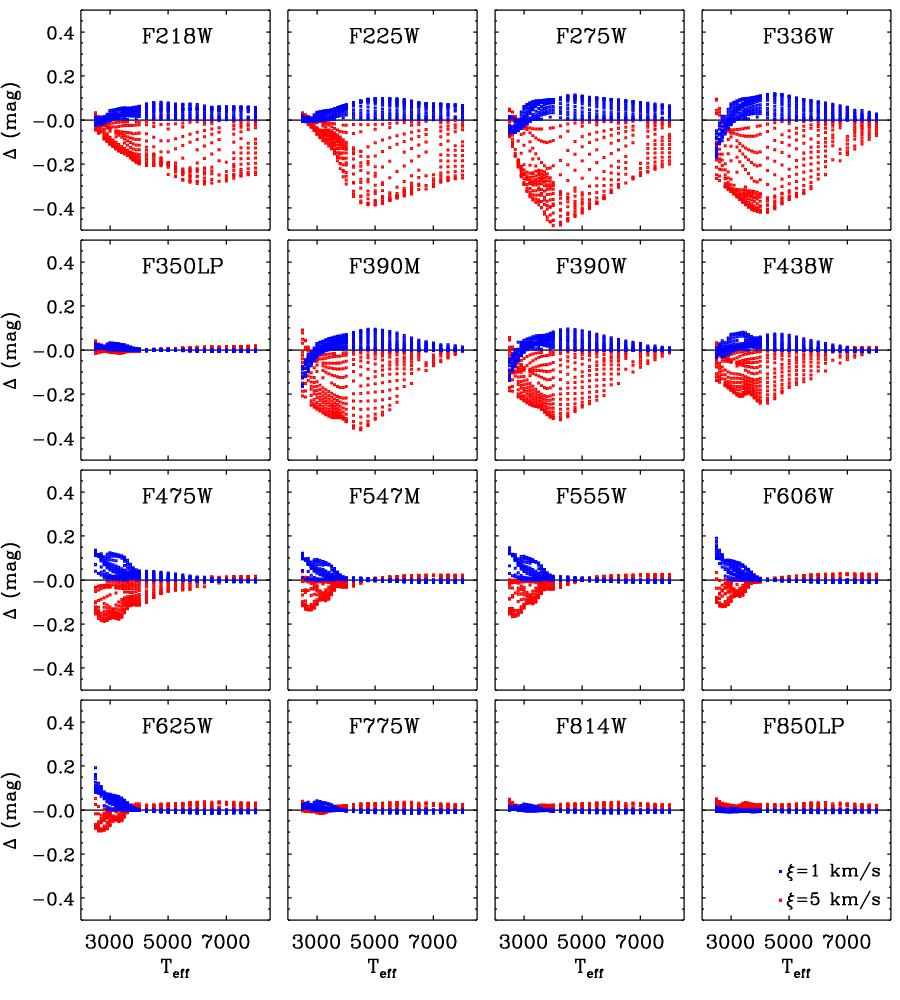}
\caption{Difference in synthetic magnitudes and $BC_\zeta$ as function of 
model $\teff$ when using a microturbulence of $1\,\kms$ (blue) or $5\,\kms$ 
(red) with respect to standard value of $\xi=2\,\kms$. The filter used is 
indicated in each panel. In all instances models with $\alpha$-standard 
composition have been used. A fixed value of $\logg=4.0$ and $2.5$ has been 
used when comparing between 1 and $2\,\kms$ and 5 and $2\,\kms$, 
respectively.}\label{f:microteff}
\end{center}
\end{figure*}

\section{MARCS synthetic library}\label{s:lib}

Magnitudes and colours have been computed using the MARCS grid of 
synthetic stellar 
spectra\footnote{http://marcs.astro.uu.se}, which is available for 
$2500 \leqslant \teff (\rm{K}) \leqslant 8000$ (in steps of $100$~K below 
$4000$~K, and $250$~K above this limit), $-0.5 \leqslant \logg 
\leqslant 5.5$ (in steps of $0.5$~dex) and different chemical compositions 
\citep{g08}. In all instances, a microturbulence $\xi=2\,\kms$ has been
adopted: we discuss later the effect of changing this parameter. 
\begin{figure*}
\begin{center}
\includegraphics[width=0.8\textwidth]{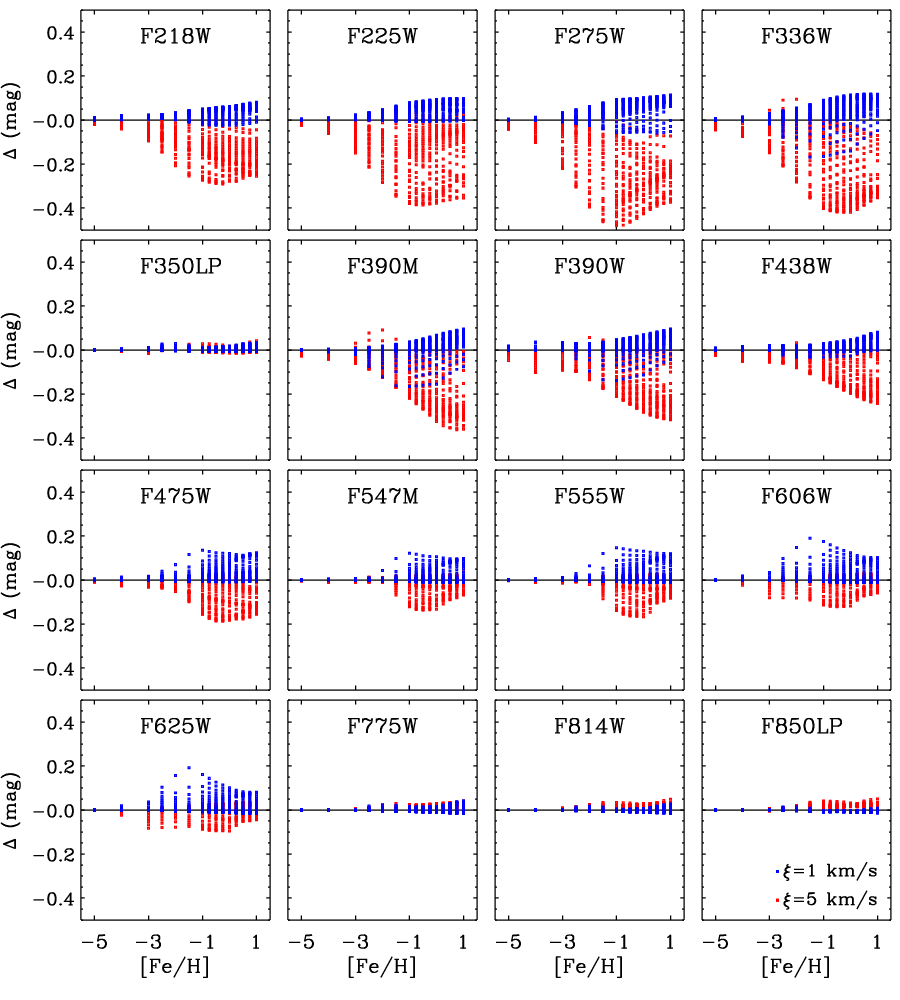}
\caption{Same as Figure \ref{f:microteff}, but as function of 
$\feh$.}\label{f:microfeh}
\end{center}
\end{figure*}

The reference solar abundance mixture is that of \cite{gas07}. All radiative 
transfer is treated assuming local thermodynamic equilibrium. The bulk of the
synthetic colours has been computed using MARCS model fluxes assuming the
so-called ``standard chemical composition" ($\alpha$-standard); i.e., a
metals mixture that reflects the values of $\aFe$ typically observed in
solar neighbourhood stars.  To be specific, $\aFe=0.4$ for $\feh
\leqslant -1.0$, $\aFe=0.0$ for $\feh \geqslant 0.0$, and $\aFe$ decreasing 
linearly from $0.4$ to $0.0$ for $-1.0 \leqslant \feh \leqslant 0.0$. 
While this $\alpha$-enrichment is broadly representative of field stars in
the Galactic disc and halo \citep[e.g.,][]{edvardsson93,nissen94,fuhrmann08},
different abundance ratios can be more appropriate for other stellar
populations, where the trend of $\aFe$ versus metallicity depends primarily
on the initial mass function and the star-formation rate 
\cite[e.g.,][]{tin79,mcwilliam97,tolstoy}. 
Models with different $\aFe$ are available from the MARCS website and we have
generated additional synthetic colours for the following cases: 
{\it i)} $\alpha$-enhanced (i.e.~$\aFe=0.4$ above $\feh=-1.0$), 
{\it ii)} $\alpha$-poor (i.e~$\aFe=0.0$ below $\feh=0.0$) and {\it iii)} 
$\alpha$-negative (i.e~$\aFe=-0.4$ from super-solar to metal-poor). The MARCS 
library covers the range $-5.0 \leqslant \feh \leqslant 1.0$, although for 
certain combinations of stellar parameters the grid contains relatively few
models.  As discussed in Appendix \ref{app}, computer programs have been 
provided
to generate BCs (and hence colour indices) for nearly the entire MARCS grid as 
it presently exists.

When computing synthetic magnitudes, we assume one solar radius at a
$10$~pc distance for all models, i.e. $f_\lambda=10^{-0.4 \rm{A}_\lambda}
 \left( \frac{R_\odot}{d_{10}} \right)^2 F_\lambda$.  While this choice is
arbitrary (and actually irrelevant), it has the practical advantage that
BCs can be computed very easily from synthetic magnitudes
(see Eq.~\ref{eq:bc}).  To determine the effect 
of varying the $E(B-V)$ from $0.0$ to $0.72$ on the BCs, 
we reddened each theoretical flux using the extinction law of \cite{od94} in 
the range $3030-9090$~\AA~and the one by \cite{ccm89} everywhere else. In all 
instances, we adopted $R_V=3.1$: an extension of our computations to other 
values of $R_V$ is currently in progress.

\subsection{The role of microturbulence}\label{sec:micro}

In one-dimensional model atmospheres, the microturbulence $\xi$ is a free 
parameter which is introduced to broaden spectral lines when synthesizing a 
spectrum. It conserves the width of weak lines, and it can reduce the 
saturation of strong lines, thus increasing their total absorption. Its 
physical interpretation is that of a non-thermal small-scale motion 
(compared to unit optical depth, or the mean free path of a photon), 
stemming from the convective velocity field in a stellar atmosphere. 
Insofar as our investigation is concerned, the effect of microturbulence is
to partly redistribute the flux in spectral regions that are crowded with
lines, and thus its effects are expected to be more evident towards the blue
and the ultraviolet, and in filters with smaller wavelength coverages. 

There is no unique, nor clear-cut, relation between the microturbulence and 
stellar parameters, although the available evidence points towards $\xi$ 
increasing at higher luminosities. For reference, a typical value for 
dwarfs and subgiants is around $1-1.5\,\kms$, going up to $2-2.5\,\kms$ for
stars on the red-giant branch (RGB), and possibly as high as $5\,\kms$ in 
supergiants \citep[e.g.,][]{nissen81,gray01,s4n,rs11}. 

A constant value of $\xi=2\,\kms$ is usually assumed in large grids of 
synthetic stellar spectra \citep[e.g.,][]{castelli04,bh05} and the same is true 
for the MARCS library as well. Fortunately, subsets of MARCS models that 
assume $\xi=1\,\kms$ and $5\,\kms$ are also available, which enables us to
investigate how a change in the microturbulence affects synthetic
magnitudes\footnote{It is for this reason that the colours which are obtained
by interpolating in our transformation tables for the solar values of $\logg$,
$\teff$\ and metallicity are somewhat different from those derived from the
MARCS synthetic spectrum for the Sun, which assumes $\xi=1\,\kms$.}.

Figures \ref{f:microteff}--\ref{f:microfeh} compare the magnitudes which are 
obtained using $\xi=1$ or $5\,\kms$ instead of $\xi=2\,\kms$. 
Notice that it follows from Eq.~(\ref{eq:bc0}) that these figures also 
illustrate the differences of $BC_\zeta$ as a function of $\teff$\ and [Fe/H].
We opted to use the WFC3 optical filters for this demonstration, as they are 
representative of most of the colours computed in this work, and they provide
a rich diversity in terms of wavelength coverage and filter widths. 
For the blue and ultraviolet filters, the effects of microturbulence are 
significant at all $\teff$ values and they increase with increasing $\feh$: in
certain cases, the differences are as much as $0.1$~mag if $\xi$ is reduced
from 2 to $1\,\kms$, or up to $0.5$~mag if the microturbulence is increased to
$\xi=5\,\kms$.  While such a high value of $\xi$ might not be representative 
of the values commonly encountered in stars, it provides a very instructive 
example
of how a fudge factor such as the microturbulence can impact synthetic colours.
The only filter essentially unaffected by a change of $\xi$ is $F350LP$, for 
which any flux redistribution occurs essentially within its very large bandpass 
($\simeq 476 {\rm nm}$). At wavelengths longer than about $5000\,$\AA~and for 
$\teff \gtrsim 4000$~K, a change of $\xi$ has less pronounced effects, 
typically of the order of, or less than, a hundredth of a magnitude.  Although 
not shown in the plots, we have verified that this is true also for  
near-infrared colours (2MASS).  In fact, these variations are at about the
same level of the accuracy at which synthetic colours can be generated when
taking into account zero-point and absolute calibration uncertainties. 

The impact of microturbulence in cool stars must therefore be kept in mind 
when comparing models with observations at wavelengths roughly bluer than $V$, 
especially for objects which are known to have microturbulent velocities that
depart considerably from the standard assumption of $2\,\kms$. 
This dependence on $\xi$ clearly introduces an additional degree of freedom 
which can be avoided only by hydrodynamic simulations that treat the velocity 
field in a consistent manner \citep[e.g.,][]{steffen13}. At the present time,
the impact of three-dimensional model atmospheres on synthetic colours is still 
largely unexplored \citep{k05,k09,c09}, but hopefully the recent advent of 
hydrodynamic grids of model atmosphere \citep{beeck13,magic13,tre13,tanner13}
will soon make such studies possible.
\begin{table}
\centering
\caption{Solar absolute magnitudes, obtained using the MARCS synthetic solar 
  spectrum ($\xi=1\,\kms$, $\teff=5777$~K, $\logg=4.44$ and $\feh=0.0$) and by
  interpolating in the $\xi=2\,\kms$ grid on the assumption of the same
  physical parameters. The filters are the same as in Table
  \ref{t:pq}.}\label{t:sun}
\begin{tabular}{ccc}
\hline 
   $M_\zeta$        & Solar     &  grid \\
\hline
$UX$               & $5.569$    &  $5.615$   \\
$BX$               & $5.427$    &  $5.438$   \\
 $B$               & $5.444$    &  $5.455$   \\
 $V$               & $4.823$    &  $4.818$   \\
 $R_C$             & $4.469$    &  $4.459$   \\
 $I_C$             & $4.129$    &  $4.120$   \\
 $U$               & $5.584$    &  $5.631$   \\
 $B$               & $5.435$    &  $5.446$   \\
 $V$               & $4.814$    &  $4.808$   \\
 $R_C$             & $4.453$    &  $4.444$   \\
 $I_C$             & $4.126$    &  $4.118$   \\
 $J$               & $3.650$    &  $3.643$   \\
 $H$               & $3.362$    &  $3.359$   \\
 $K_S$             & $3.277$    &  $3.274$   \\
 $u$               & $6.430$    &  $6.479$   \\
 $g$               & $5.075$    &  $5.079$   \\
 $r$               & $4.649$    &  $4.640$   \\
 $i$               & $4.535$    &  $4.526$   \\
 $z$               & $4.508$    &  $4.500$   \\
 $F435W_{\rm{ACS}}$ & $5.442$    &  $5.454$   \\
 $F475W_{\rm{ACS}}$ & $5.169$    &  $5.172$   \\
 $F555W_{\rm{ACS}}$ & $4.832$    &  $4.827$   \\
 $F606W_{\rm{ACS}}$ & $4.631$    &  $4.623$   \\
 $F814W_{\rm{ACS}}$ & $4.109$    &  $4.100$   \\
 $F218W$           & $9.220$    &  $9.263$   \\
 $F225W$           & $8.446$    &  $8.517$    \\
 $F275W$           & $6.986$    &  $7.063$    \\
 $F336W$           & $5.465$    &  $5.523$    \\
 $F350LP$          & $4.751$    &  $4.748$    \\
 $F390M$           & $5.987$    &  $6.018$    \\
 $F390W$           & $5.611$    &  $5.638$    \\
 $F438W$           & $5.447$    &  $5.460$    \\
 $F475W$           & $5.146$    &  $5.149$    \\
 $F547M$           & $4.800$    &  $4.794$    \\
 $F555W$           & $4.868$    &  $4.864$    \\
 $F606W$           & $4.639$    &  $4.632$    \\
 $F625W$           & $4.493$    &  $4.484$    \\
 $F775W$           & $4.156$    &  $4.147$    \\
 $F814W$           & $4.115$    &  $4.106$    \\
 $F850LP$          & $4.001$    &  $3.993$    \\
 $F098M$           & $3.956$    &  $3.949$    \\
 $F110W$           & $3.787$    &  $3.780$    \\
 $F125W$           & $3.662$    &  $3.656$    \\
 $F140W$           & $3.520$    &  $3.515$    \\
 $F160W$           & $3.393$    &  $3.390$    \\
\hline
\end{tabular}
\begin{minipage}{1\textwidth}
HST magnitudes are all in the {\tt VEGA} system.
\end{minipage}
\end{table}

\section{Comparisons with observations}
\label{sec:obs}

The testing of synthetic fluxes is normally done by comparing the observed and
modelled spectral energy distributions for a range of wavelengths and spectral
types \citep[e.g.,][]{edvardsson08,boh10,bes11} or by comparing synthetic and
observed colours in different bands and for different values of $\teff$,
$\logg$ and $\feh$ \citep[e.g.,][]{b98,onehag09}. Here we follow the latter
approach.  First, we compare synthetic colours in different filters with
observations of field stars that have well determined physical parameters and
magnitudes, as well as with empirical relations that link colours to stellar
parameters. Then, we turn our attention to star clusters, which enable us to
test the consistency of the fits of isochrones to the observed photometry 
on several different colour-magnitude planes.
\begin{figure}
\begin{center}
\includegraphics[width=0.45\textwidth]{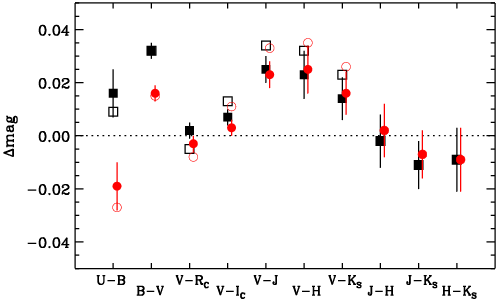}
\caption{Solar synthetic colours compared to empirical determinations from 
Ram\'irez et al.~(2012) and Casagrande et al.~(2012). Squares are 
obtained using the MARCS solar flux (which has $\xi=1\,\kms$), while circles 
(slightly shifted on the abscissa for clarity) are from interpolations 
in the $\xi=2\,\kms$ MARCS grid at the solar values of $\teff$, $\logg$, and 
$\feh$. Only the error bars from the above empirical determinations have been 
attached to the synthetic magnitudes. For the optical filters, filled and open 
symbols refer to the {\tt ubvri90} and {\tt ubvri12} transformations, 
respectively.}\label{f:sun}
\end{center}
\end{figure}

Readers should bear in mind the caveats of generating synthetic photometry that
have been discussed in Section \ref{sec:sp}, as well as the difficulty of
standardizing photometric observations (ground-based, in particular) to better
than about $0.01$~mag \citep[e.g.,][]{st05}.  Therefore, when comparing
synthetic and observed colours, agreement to within $0.01-0.02$~mag should
usually be regarded as excellent. 

\subsection{The Sun and other field stars}
\label{subsec:field}

Arguably, the Sun is the star with the best-known parameters, and it is the
most important benchmark in stellar astrophysics.  However, for obvious
reasons, it cannot be observed with the same instruments and under the same
conditions as more distant stars, thus making it virtually impossible to tie
its colours to other photometric observations \citep[e.g.,][]{sk57}. 
Recently, the use of solar analogs and twins has overcome 
this limitation, and reliable colours are now available in the Johnson-Cousins 
and 2MASS system \citep{r12,c12}. Figure \ref{f:sun} compares these 
measurements with the colour indices obtained using the MARCS solar synthetic 
flux (open circles). Differences are usually well within $0.02$~mag, which is
approximately the conjugated uncertainty of the observed and synthetic 
colours, and hence the accuracy at which our colours are testable. Also shown 
in Figure \ref{f:sun} is a similar comparison, but of colours which are 
obtained by interpolating in 
the $\xi=2\,\kms$ grid at the solar $\teff$, $\logg$, and $\feh$.  The
differences with respect to those based on the solar synthetic spectrum are
negligible longward of the $V$ band, whereas at wavelengths bluer than the 
$B$ band they are more substantial (in accordance with the effects of 
microturbulance discussed in Section \ref{sec:micro}).

\begin{figure*}
\begin{center}
\includegraphics[width=0.99\textwidth]{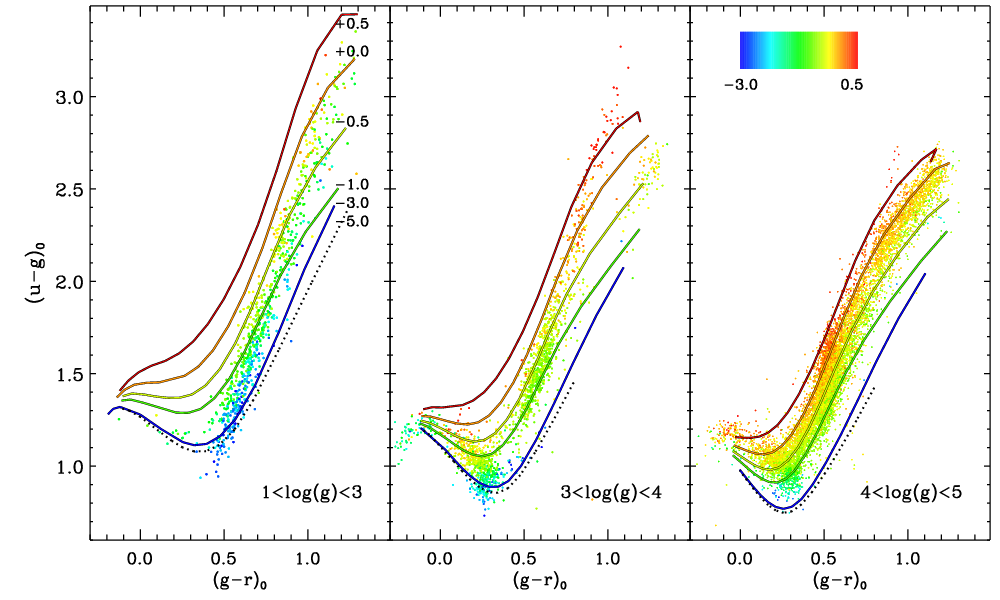}
\caption{Comparison between observed and synthetic $ugriz$ colours 
($\alpha$-standard, continuous lines) for a sample of about $13500$ stars in 
SEGUE DR8. Stars are dereddened using $E(B-V)$ values provided by SEGUE, sorted 
into panels according to surface gravity, and colour coded by metallicity 
($\logg$ and $\feh$ from the SSPP pipeline). MARCS ($\alpha$-standard) colours 
are shown for fixed $\feh$ from $-5.0$ to $+0.5$ (as indicated) and $\logg=2$ 
(left), $3.5$ (centre) and $4.5$ (right-panel), for $\teff \ge 
4000$~K.}\label{f:ugriz}
\end{center}
\end{figure*}

Figure \ref{f:ugriz} shows the comparison between the observed and 
synthetic $ugriz$ colours for a sample of about $13500$ stars in the SEGUE DR8 
catalogue \citep{ya09}, flagged to have good photometric quality in all bands, 
and to cover a broad range of stellar parameters as determined from the SSPP 
pipeline \citep[][and references therein]{lee11}. MARCS synthetic colours are 
clearly quite successful in reproducing the observed trends on the different
colour planes as function of both $\feh$ and $\logg$, especially given that
the SSPP values have their own measurement uncertainties.  Also worth pointing
out is the decreasing sensitivity of $ugriz$ broad-band colours to $\feh$ in
the direction of reduced metallicities: note the reduced separations of the
loci between $-1 \geqslant \feh \geqslant -3$, and even more so between
$-3 \geqslant \feh \geqslant -5$.

\begin{figure*}
\begin{center}
\includegraphics[width=0.9\textwidth]{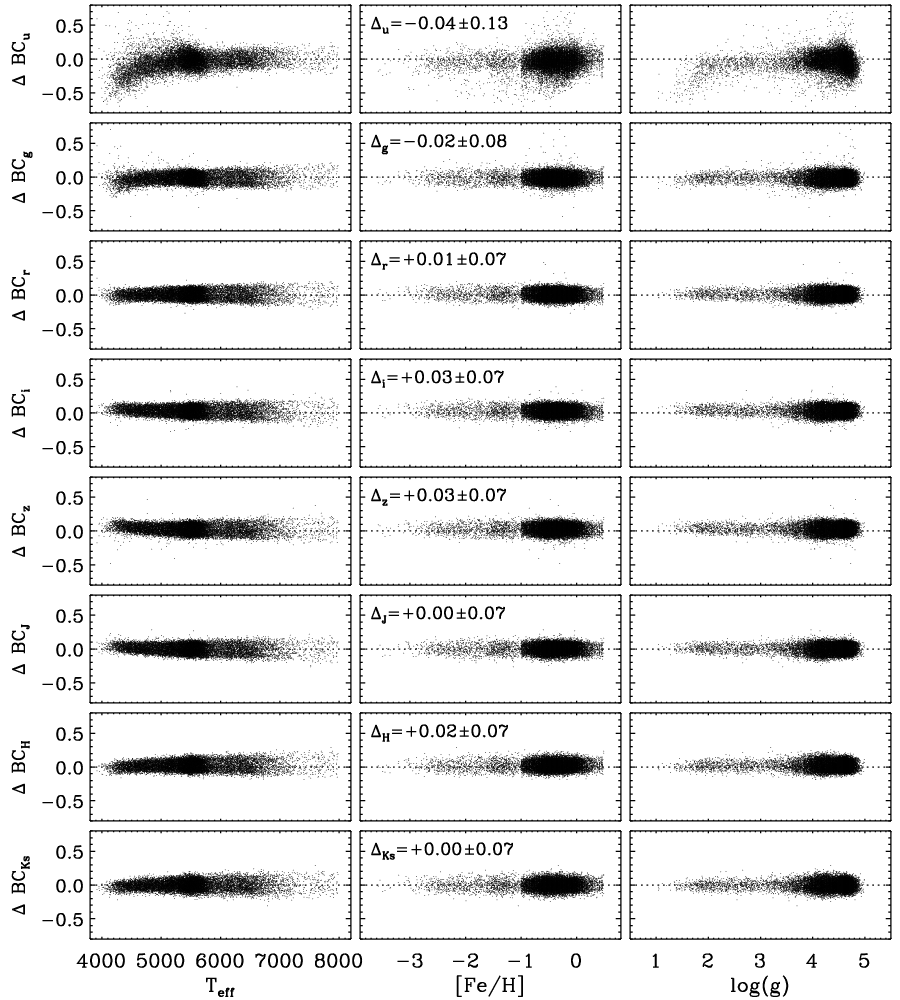}
\caption{Comparison between empirical and MARCS ($\alpha$-standard) BCs for 
stars in the previous Figure. For each band, the mean difference and the 
standard deviation are reported in the central panel.}\label{f:BC}
\end{center}
\end{figure*}

For all of the stars in Figure \ref{f:ugriz}, we also determine $\teff$ and 
$\fbol$ using the Infrared Flux Method (IRFM) described in \cite{c10}. 
Briefly, multi-band optical $ugriz$ and infrared $JHK_S$ photometry is used to 
recover the bolometric flux of each star, from which its effective temperature 
can then be readily obtained. The method is to a large extent empirical: it 
depends very mildly on the input $\feh$ and $\logg$ of each star (adopted 
from SSPP in our case) and it uses so many photometric bands that it relies 
very little on synthetic flux libraries --- a claim that has been extensively 
tested in the literature \citep[e.g.,][]{blackwell80,alonso95,c06}. In Figure 
\ref{f:BC},
we compare, for each star, the empirical BCs derived in the
$ugriz$ and $JHK_S$ systems when using $\fbol$ from the IRFM, or as obtained
instead by interpolating in our grids of MARCS synthetic photometry at the 
$\feh$, $\log$ (from SSPP) and $\teff$ (from the IRFM) of each star. 

This comparison allows us to easily quantify the performance of the MARCS 
synthetic fluxes in predicting broad-band colours on a star-by-star basis. In 
fact, it follows from Eq.~(\ref{eq:bc0}) that a difference of
$\Delta_\zeta$~mag in a bolometric correction corresponds to a fractional
uncertainty of $10^{-0.4\Delta_\zeta}$ in the derived bolometric flux.  The
BCs calculated from MARCS models usually agree to within
$0.02 \pm 0.07$~mag with the values determined empirically from the IRFM
for large ranges of $\teff$, $\feh$, and $\log$.  Only in the $u$ band
is the uncertainty larger, and it shows systematic deviations towards the
lowest surface gravities. The above numbers translate into a typical
uncertainty of about $2 \pm 7$ per cent in the bolometric fluxes that are
obtained from the MARCS tables. It should also be kept in mind that $\fbol$
values from the IRFM typically have errors of a few percent.  Overall, this
implies that the grid of MARCS synthetic broad-band colours reported here
can be used to reconstruct the bolometric flux of a star of known $\teff$,
$\logg$, and $\feh$ to within an accuracy of a few per cent, and 
with a precision comparable to what can be achieved with empirical 
approaches. Also, the comparisons presented in Fig.~\ref{f:BC} for a wide range
of physical parameters in the $ugrizJHK_S$ system is representative of the
wavelength domain explored in this work, and of the parameter space covered
by the MARCS library as a whole.

\subsection{Colour--$\teff$--metallicity relations}
\label{subsec:cmt}

Empirical colour--$\teff$--metallicity relations offer another way to test how 
well the interplay among these quantities is reproduced by synthetic flux
models --- before we ``attach'' them to stellar isochrones (the subject of
Section \ref{subsec:starcl}.) Using the routines described in Appendix 
\ref{app}, we generated BCs in the $BVI_CJK_S$ system (from
which colour indices in any combination of the above filters can be derived)
for two metallicities, $\feh=0.0$ and $-2.5$, and $\teff$ values from $4500$~K
to $6500$~K, in steps of $100$~K, at two fixed values of $\logg=4.0$ and $4.5$.
These choices broadly encompass the properties of most of the subgiants and
dwarf stars in the Galaxy.

\begin{figure}
\begin{center}
\includegraphics[width=0.45\textwidth]{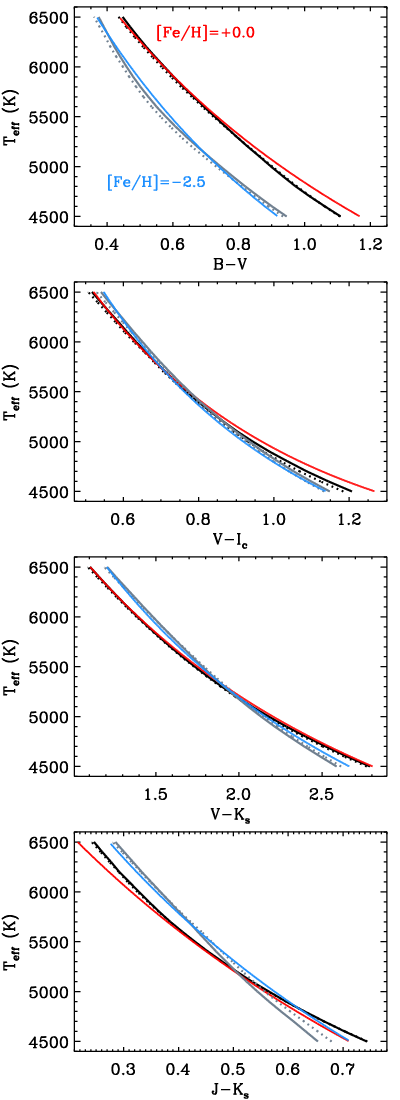}
\caption{Comparison between the empirical colour--$\teff$--metallicity 
relations of \citealt{c10} ($\feh=0.0$ red, $\feh=-2.5$ blue), and those 
obtained from the MARCS ($\alpha$-standard) fluxes at the same metallicity 
($\feh=0.0$ black, $\feh=-2.5$ grey) for two different gravities, $\logg=4.0$ 
(dotted line) and $4.5$ (continuous line).}\label{f:colorteff}
\end{center}
\end{figure}

For a fixed $\logg$ we can then produce synthetic colour--$\teff$--metallicity
relations to be compared with the polynomials given by \cite{c10}: such
comparisons are presented in Figure \ref{f:colorteff}.  The empirical relations
have a ``built-in''surface gravity dependence as they are valid for 
MS and subgiant stars.  As a consequence, we chose to generate
synthetic colours for values of $\teff, \feh$, and $\logg$ that are
representative of the majority of the stars on which the empirical polynomials
are based to do a better sampling of the parameter space which contains the
observed stars. Figure \ref{f:colorteff} shows the effect of using two different
gravities mainly to demonstrate that the exact choice of $\logg$ is not the
dominant factor in this comparison. 

It is also worth noticing that the absolute calibration, zero-points, and
filter transmission curves used in \cite{c10} to derive the empirical relations 
closely match those used to produce synthetic MARCS BCs in 
this study (see Section \ref{subsec:jc2m}). The comparisons performed here are 
thus largely free from any systematic biases that the above choices could
introduce, and thus allows us to directly assess the degree to which the
synthetic fluxes are able to match empirical relations. Overall, synthetic
colours reproduce the main trends that are seen in empirical relations, as
well as the crossing-over of the fiducial sequences for different values of 
$\feh$ in certain colour indices. The agreement is particularly good 
around $5500$~K in all of the colour indices which are examined in Figure 
\ref{f:colorteff}.  The largest offsets are found for the $B-V$ colour index
at lower temperatures and higher metallicities --- which is not too surprising,
considering the increased crowding of molecular lines in metal-rich dwarfs.
Similarly, the predicted $J-K_S$ colours that we have derived from hot 
metal-rich
($\teff \gtrsim 6000$~K) and cool metal-poor ($\teff \lesssim 5200$~K) models
increasingly depart from the empirical relations. In this case, however, it
should be kept in mind that the scatter around the empirical $J-K_S$ relation
is quite large, partly stemming from its limited colour baseline, and that
it seems to underestimate $\teff$ above $6000$~K \citep{pam12}. Nevertheless,
all things considered, the comparisons given in Figure \ref{f:colorteff}
validate the MARCS effective temperature scale (for dwarfs and subgiants,
in particular) when using indices such as $V-I_C$ or $V-K_S$, which have been
shown to agree with the empirical \cite{c10} scale, typically to much better
than $100$~K.

\subsection{Open and Globular Clusters}
\label{subsec:starcl}

Comparisons of computed isochrones with the observed colour-magnitude diagrams
(CMDs) of star clusters provide valuable tests of the consistency of synthetic
colour--$\teff$ relations because the same interpretation of the 
data should 
be obtained on all CMDs.  This is not to say that the models must provide
perfect reproductions of the observed morphologies, but rather that whatever
discrepancies exist between theory and observations in one CMD should be
consistent with those that are apparent on other colour-magnitude planes. When 
that is not the case, this information can be used to pinpoint where 
inaccuracies occur in synthetic fluxes. 
To illustrate this, we have opted to consider the extensively studied globular
cluster M\,5, which has [Fe/H] $= -1.33$ according to \citet{cbg09}, the old,
super-metal-rich open cluster NGC\,6791 ([Fe/H] $\approx +0.3$; see
\citealt{bvb12}), and the near solar-metallicity open cluster M\,67.  (It
should be appreciated that some discrepancies between predicted and observed
CMDs in an absolute sense will generally be present due to the uncertainties
associated with, e.g., the treatment of convection and the atmospheric boundary
conditions in stellar models, as well as those connected with the photometric
data and such basic cluster properties as the distance, reddening, and chemical
abundances.)

In what follows, observations of these three systems are compared with the
latest Victoria-Regina isochrones (\citealt{vbf14}) on several different CMDs.
Rather than correcting the photometry for the effects of reddening and distance
to obtain the intrinsic colours and absolute magnitudes, the model luminosities
and temperatures have been converted to the {\it observed} magnitudes and 
colours using the reddening-adjusted transformations presented in this 
investigation.  To accomplish this, values of $(m-M)_0$ (true distance moduli)
that agree well with published estimates have been assumed. In all instances,
well-supported determinations of the metallicity have been adopted, along with
$E(B-V)$ values from \citet[hereafter SF11]{sf11}, except in the case of
NGC$\,$6791.  For this cluster, a slightly higher reddening, by $\approx 0.02$
mag, appears to result in improved consistency over most of the colour planes
that have been considered (see also discussion in Section \ref{sec:morpho},
though it may instead be the assumed [Fe/H] value
that is not quite right).  Such issues are not of particular importance for
the present analysis, however, as our primary goal is to study how the 
{\it differences} between the isochrones and the cluster CMDs vary with the
particular colour index that has been plotted.

\subsubsection{M$\,$5 (NGC$\,$5904)}
\label{subsubsec:m5}  
   
Thanks to the efforts of P.~B.~Stetson, very high-quality, homogeneous 
Johnson-Cousins $BV(RI)_C$ photometry is available for relatively large 
samples of
stars in many open and globular clusters (\citealt{st05}).  For a few of them,
including M$\,$5, $U$-band data may also be obtained from his ``Photometric
Standard Fields" archive at the Canadian Astronomy Data Centre.\footnote{See
www1.cadc-ccda.hia-iha.nrc-cnrc.gc.ca/community/STETSON/}  
Indeed, his
observations yield exceedingly tight, well-defined CMDs for M$\,$5 from the
tip of the RGB to $\sim 2$ mag below the turn-off (TO).  A
further reason for choosing to use this system to examine the colour
transformations for metal-poor stars is that it is nearly unreddened [$E(B-V) =
0.032$ according to SF11].
\begin{figure*}
\begin{center}
\includegraphics[width=0.98\textwidth]{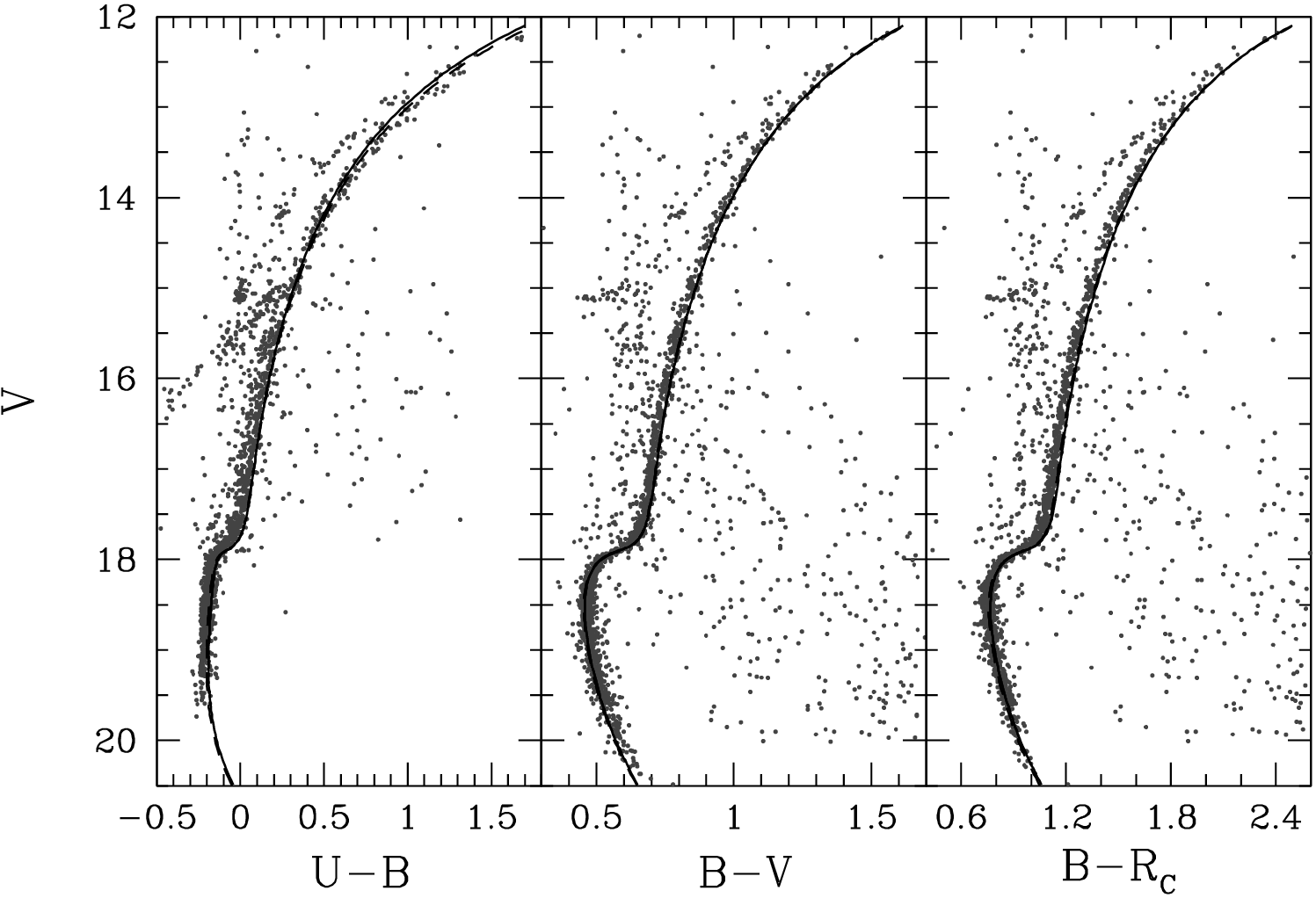}
\caption{Overlay of an 11.75 Gyr isochrone for [Fe/H] $= -1.33$, [$\alpha$/Fe]
$=0.4$, and $Y = 0.25$ onto three CMDs of M$\,$5 (see the text for the sources
of the stellar models and the photometry).  The solid and dashed curves assume
the reddening-adjusted {\tt ubvri12} and {\tt ubvri90} colour transformations,
respectively, for $E(B-V) = 0.032$ (from SF11).  The true distance modulus of
M$\,$5 has been assumed to be $(m-M)_0 = 14.35$.}
\label{fig:fig14}
\end{center}
\end{figure*}

Figure~\ref{fig:fig14} plots the measured $V$ magnitudes of the stars in the
observed fields of M$\,$5 as a function of their $U-B$, $B-V$, and $B-R_C$ 
colours. The solid curve in each panel is an 11.75 Gyr \citep{vbl13} isochrone 
for [Fe/H] $= -1.33$, [$\alpha$/Fe] $= 0.4$, and a helium abundance 
$Y = 0.25$.  To place it on the
observed planes, the so-called {\tt ubvri12} transformations 
for $E(B-V) = 0.032$ were used, together with $(m-M)_0 = 14.35$.  Dashed loci
represent the same isochrone, except that the {\tt ubvri90} transformations 
have been employed: differences between them and the solid curves are 
apparent only in the left-hand panel.  Historically, synthetic $U-B$ colours 
have not been very reliable, but those based on the latest MARCS model
atmospheres appear to be quite good (if the metallicity and distance that we 
have assumed for M$\,$5 are accurate). Indeed, if the predicted $U-B$ colours 
were adjusted to the blue by
only 0.02--0.03 mag, the isochrone would provide reasonably consistent fits to
the observations on all three of the CMDs that are shown in 
Fig.~\ref{fig:fig14}.
That is, independently of the colour that is considered, the isochrone would
match the TO photometry satisfactorily and be somewhat too red along the
RGB --- except at $V \lta 15$ on the $(U-B,\,V)$-diagram, which suggests that
the $U-B$ colours become less trustworthy at low gravities and/or cool
temperatures. This is also in agreement to what was shown in Figure 
\ref{f:BC} for the $u$ filter.
\begin{figure}
\begin{center}
\includegraphics[width=0.48\textwidth]{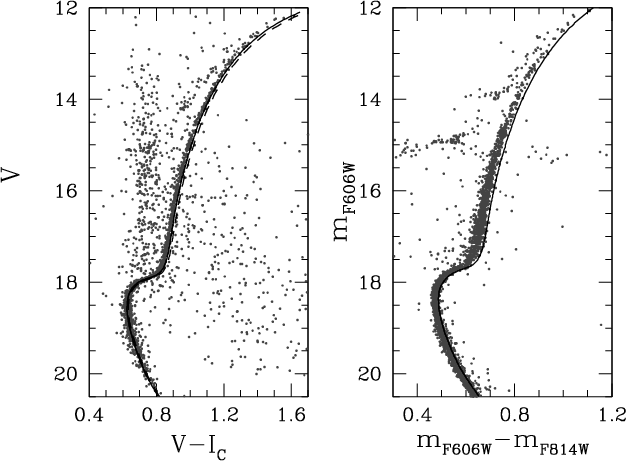}
\caption{The left-hand panel extends the plots in the previous figure to the
$(V-I_C,\,V)$-diagram.  In the right-hand panel, HST ACS data for M$\,$5 from 
\citet{sbc07} have been plotted, and the isochrone has been converted to the 
observed plane using the relevant transformations in the {\tt VEGA-MAG} system 
(see Appendix \ref{app}).} 
\label{fig:fig15}
\end{center}
\end{figure} 
 
The left-hand panel of Figure~\ref{fig:fig15} shows that the same consistency
is found on the $(V-I_C,\,V)$-plane. Interestingly, this plot indicates a slight
preference for the {\tt ubvri12} transformations to $V-I_C$\ over those that we
have labelled {\tt ubvri90}, though this result will depend on the model
$\teff$ scale, which is subject to many uncertainties.  On the other hand, one
would not expect to find such good agreement between the predicted and observed
RGB morphologies unless the temperatures along the isochrones are quite close
to those implied by the MARCS colour--$\teff$ relations. As illustrated in the
right-hand panel, very similar results are obtained when the same isochrone is
compared with the HST ACS observations of M$\,$5 that
were obtained by \citet{sbc07}.  However, in this case, the isochrone does not
match the observations quite as well as in the left-hand panel: to achieve a
fully consistent fit, the models would need to be adjusted to the blue by 
0.01--0.015 mag.  It is not clear whether this zero-point correction should be
made to the colour transformations or to the photometric data (from either 
source).

\begin{figure}
\begin{center}
\includegraphics[width=0.48\textwidth]{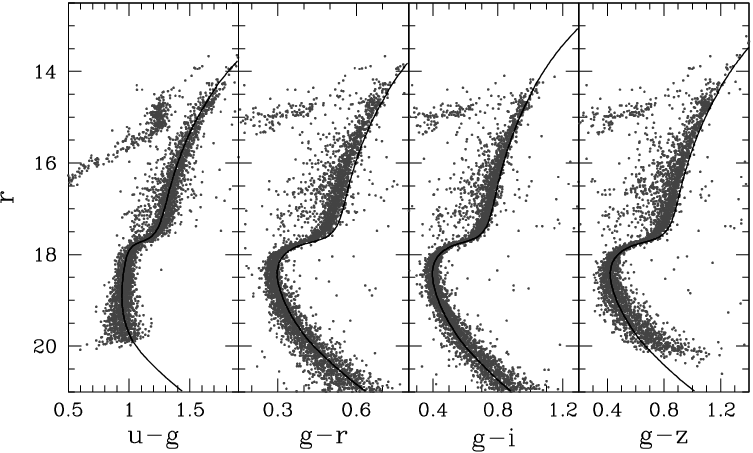}
\caption{As in Fig.~\ref{fig:fig14}, except that the isochrone has
been overlaid onto four CMDs derived from the Sloan ($ugriz$) photometry of
M$\,$5 obtained by \citet{ajc08}.}
\label{fig:fig16}
\end{center}
\end{figure}

The availability of Sloan ($ugriz$) observations of M$\,$5 by \citet{ajc08}
gives us the opportunity to extend our comparisons to this photometric system.
Four of the many possible CMDs that could be constructed from their observations
are shown in Figure~\ref{fig:fig16}, along with the same isochrone that has been
plotted in the two previous figures, on the assumption of the same reddening
and true distance modulus, etc.  The fits to the data on the $g-i$ and $g-z$
colour planes closely resemble the fits to the $BV(RI)_C$ observations discussed
above, which implies satisfactory consistency between them.  However, as in the
case of the ACS photometry of M$\,$5, the isochrone lies $\sim 0.02$ to the red
of the cluster MS and TO on the $(g-r,\,r)$-diagram, while it reproduces the
observed $u-g$ colours (in the mean) for both the MS and and RGB nearly
perfectly.  To try to identify whether the $g$ or $r$ magnitudes are primarily
responsible for the offset in the $g-r$ colours, we produced a plot on the
$(r-i,\,r)$-plane (not shown).  It revealed that the isochrone is too
blue relative to the observed TO by about 0.02 mag, from which we conclude
that the difficulty with the $g-r$ colours is mainly due to a small problem with
the $r$ magnitudes.  Whether this should be attributed to our transformations
or to the observations, or both, is not known.

To improve our understanding of star clusters and of the photometric systems
that are used, it is clearly advantageous to obtain observations through as
many filters as possible.  Encouragingly, most of the plots that we have
produced for M$\,$5 contain basically the same fit of the selected isochrone
to the observations.  In other words, the models generally provide a consistent
interpretation of the data, regardless of the particular CMD that is considered.
However, we did find a few ``exceptions to the rule", which is to be expected
given the difficulty of establishing the zero-points of synthetic or observed
photometry to within $\sim 0.01$--0.02 mag and the uncertainties associated
with current model atmospheres and synthetic spectra.  In fact, the small colour
offsets that we found on some colour-magnitude planes may also be a reflection
of the assumed metal abundances; i.e., perhaps improved overall consistency
would have been obtained had our analysis used stellar models that assumed a
different mixture of the heavy elements.  Although it is beyond the scope of
the present investigation to do so, this possibility should be fully explored. 

\subsubsection{NGC$\,$6791}
\label{subsubsec:n6791}

\cite{bbg11,bvb12} have carried out the most exhaustive studies
of NGC$\,$6791 to date, finding an age of 8--8.5 Gyr if $Y \approx 0.30$ and
[Fe/H] $= 0.30$--0.35, when fits of stellar models to the mass-radius diagrams
of the binaries known as V18 and V20 are taken into account.  Their results
are based predominately on isochrones that assume the solar mix of the heavy
elements as reported by \citet{gs98}, suitably scaled to the [Fe/H] values of
interest.  For the present work, which uses the latest Victoria-Regina
isochrones (\citealt{vbf14}), the updated solar abundances by \citet{ags09}
provide the reference metals mixture.  According to VandenBerg et al., their
models require $Y \approx 0.28$ to satisfy the same binary constraints if
NGC$\,$6791 has [Fe/H] $= 0.30$ --- and we have therefore adopted these
estimates of the helium and metallicity.
\begin{figure}
\begin{center}
\includegraphics[width=0.48\textwidth]{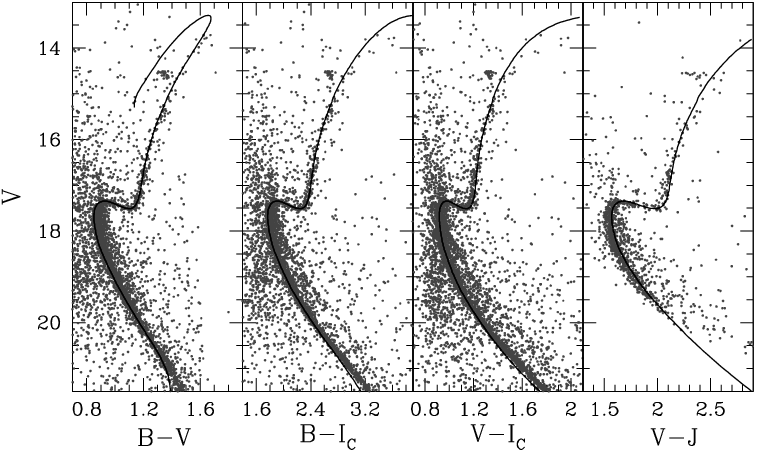}
\caption{Overlay of an 8.5 Gyr isochrone for [Fe/H] $= +0.30$, [$\alpha$/Fe]
$= 0.0$, and $Y = 0.28$ onto four CMDs of NGC$\,$6791 (see the text for the
sources of the stellar models and the observations).  The solid curves assume
the reddening-adjusted {\tt ubvri12} and the {\tt SDSS} $J$ transformations for
$E(B-V) = 0.16$.  The true distance modulus of NGC$\,$6791 has been assumed to
be $(m-M)_0 = 13.05$.}
\label{fig:fig17}
\end{center}
\end{figure}

Unfortunately, the reddening of NGC$\,$6791 has been particularly hard to pin
down.  The value of $E(B-V) = 0.155$ from the original \citet{sfd98} dust maps
has been revised downwards to 0.133 mag by SF11.  Moreover, as discussed by
\citet{bvb12}, who concluded that the current best estimate of the reddening is
$0.14 \pm 0.02$, it appears to be difficult to obtain consistent fits to the
$(B-V,\,V)$- and $(V-I_C,\,V)$-diagrams unless a small systematic shift is 
applied to one of the predicted colours.  Part of the difficulty, as noted by 
them, is that the value of $E(V-I_C)/E(B-V)$ appears to be uncertain (e.g., 
\citealt[their
Tables 11--13]{hsv12}).  As it turns out, it {\it is} possible to obtain
nearly indistinguishable fits of isochrones to observations of NGC$\,$6791 on 
many different CMDs (using our predictions for the effects of reddening on the
synthetic magnitudes and colours) if the cluster has $E(B-V) = 0.16$.
 
\begin{figure}
\begin{center}
\includegraphics[width=0.48\textwidth]{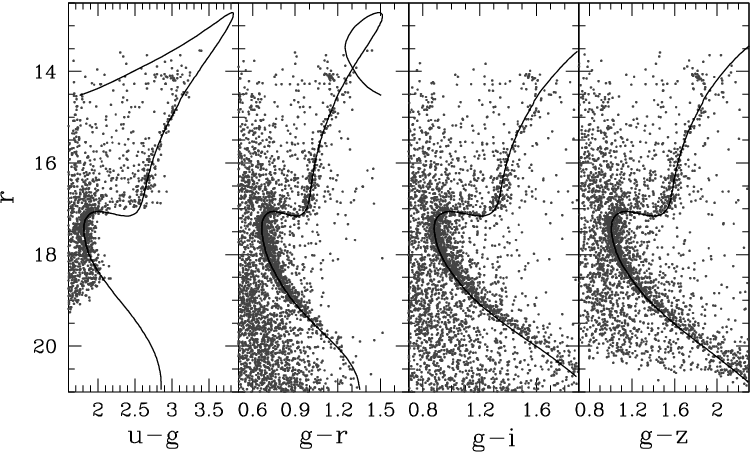}
\caption{As in the previous figure, except that the isochrone has been 
overlaid onto four CMDs derived from the Sloan ($ugriz$) photometry of 
NGC$\,$6791 obtained by \citet{ajc08}.}
\label{fig:fig18}
\end{center}
\end{figure}

This is shown in Figures~\ref{fig:fig17} and~\ref{fig:fig18} where, in turn,
an 8.5 Gyr isochrone for the aforementioned chemical abundances has been
overlaid onto a number of CMDs derived from $BVI_CJ$ and $ugriz$ photometry of
NGC$\,$6791.  All fits assume $(m-M)_0 = 13.05$.  The $BVI_C$ photometry that we
have used is contained in the {\tt NGC6791.KFB} and {\tt NGC6791.COR} files
that can be downloaded from the ``Homogeneous Photometry" link at the same CADC
website address (see footnote 14) that was used to obtain the M$\,$5 data.
Although $U$-band observations are not available, we have been able to extend
our comparisons to the near-infrared by adding a panel that plots the Johnson
$V$, 2MASS $J$ photometry from \citet{bsv10}.  For whatever reason, the
isochrone lies along the red edge of the distribution of MS stars on the 
$(V-J,\,V)$-diagram, while it coincides closely with the blue edge
in the other panels, as well as in at least three of the four CMDs shown in
Fig.~\ref{fig:fig18}.  [The isochrone appears to match the mean MS fiducial
on the $(u-g,\,r)$-plane, but the TO is close to the limit of the $u$
photometry.]    As in the case of M$\,$5, the Sloan photometry of NGC$\,$6791
that we have used has been taken from \citet{ajc08}.

Even in an absolute sense, the isochrone clearly reproduces the observed CMDs
of NGC$\,$6791 rather well from 3--4 mag below the TO to approximately the
luminosity of the horizontal branch clump.  At higher luminosities, the
predicted colours tend to be too blue, indicating a problem with either the
MARCS transformations or the model temperatures or both.  (For quite a thorough
discussion of this issue, reference may be made to \citealt{vbf14}.)  Although
Figs.~\ref{fig:fig17} and~\ref{fig:fig18} appear to provide compelling support
for the derived (or adopted) properties of NGC$\,$6791, we have not explored
whether equally good fits of the isochrones to the observed CMDs (and the
cluster binaries) can be obtained for different combinations of [Fe/H], the
mix of heavy elements, $Y$, $E(B-V)$, $(m-M)_0$, and age.  This would be well
worth investigating.  Note that the unusual morphology of the upper RGB portion
of the isochrone in the $(B-V,\,V)$-, $(u-g,\,r)$-, and $(g-r,\,r)$-diagrams
is consistent with the predicted behaviour of the synthetic magnitudes in the
respective filters, and are not a consequence of, say, extrapolations in the
colour transformation tables.  On the other hand, that morphology will be a
sensitive function of the model $\teff$\ scale; i.e., if the predicted RGB
were appreciably cooler or hotter, the blueward excursion of upper RGB in the
left-hand panel of Fig.~\ref{fig:fig18} and the loop in the adjacent panel 
would be accentuated or minimized.

\subsubsection{M$\,$67 (NGC$\,$2682)}
\label{subsubsec:m67}

Our emphasis thusfar has been on upper MS and more evolved stars because several
recent papers have already shown that the slope of the MS down to $M_V \sim
7$--8 in open and globular clusters and as defined by local subdwarfs, over
a wide range in [Fe/H], can be reproduced quite well by stellar models that
employ the MARCS colour transformations (see \citealt[2013, 2010]{vbf14};
\citealt{bsv10}).  Hence, for our final comparisons of 
isochrones with cluster observations, we decided to focus on the lower-MS
extension of M$\,$67, given that $ugriz$ (\citealt{ajc08}) and $BVI_CJK_S$
photometry is available for low-mass stars belonging to this system. [The
2MASS observations (\citealt{scs06}) were kindly provided to us by
A.~Dotter (see \citealt{sdk09}), while the Johnson-Cousins photometry is from
the same CADC website mentioned previously (see footnote 14).] 

\begin{figure}
\begin{center}
\includegraphics[width=0.48\textwidth]{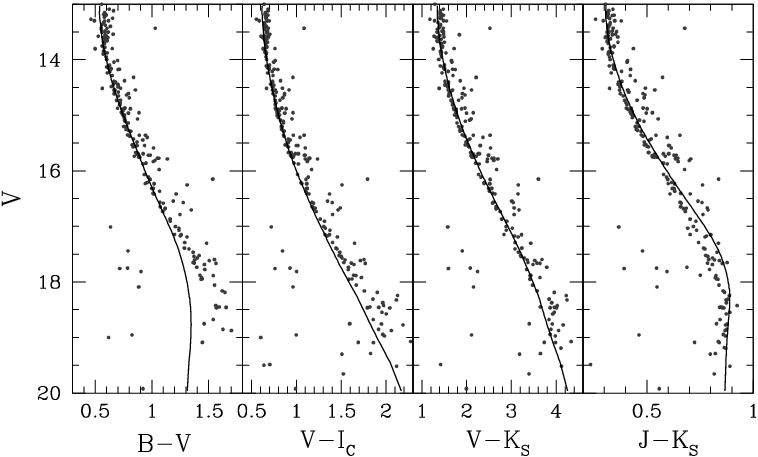}
\caption{Overlay of the lower-MS portion of a $4.3$ Gyr isochrone
for [Fe/H] $= 0.0$, [$\alpha$/Fe] $= 0.0$, and $Y = 0.255$ onto four CMDs
for MS stars in M$\,$67 (see the text for the sources of the stellar models
and the observations).  The solid curves assume the reddening-adjusted
{\tt ubvri12} and the {\tt SDSS} $JK_S$ transformations for $E(B-V) = 0.030$.
The true distance modulus of M$\,$67 has been assumed to be $(m-M)_0 = 9.60$.}
\label{fig:fig19}
\end{center}
\end{figure}

\begin{figure}
\begin{center}
\includegraphics[width=0.48\textwidth]{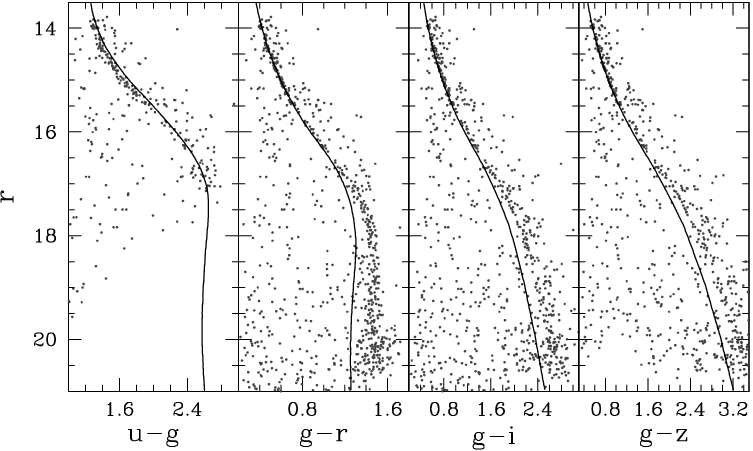}
\caption{As in the previous figure, except that the isochrone has been overlaid
onto four CMDs derived from the Sloan ($ugriz$) photometry of M$\,$67 obtained
by \citet{ajc08}.}
\label{fig:fig20}
\end{center}
\end{figure}

In Figures~\ref{fig:fig19} and~\ref{fig:fig20}, eight of the possible CMDs that
can be produced from these data have been superimposed by the MS segment of the
same 4.3 Gyr, [Fe/H] $= 0.0$ (e.g., \citealt{rsp06}, \citealt{okg11}),
$Y = 0.255$ isochrone that was compared with observations for the upper MS and
RGB stars of M$\,$67 by \citet{vbf14}.  In all cases, $E(B-V) = 0.030$ (SF11)
and $(m-M)_0 = 9.60$ has been assumed.  Except for the $J-K_S$ and $u-g$ colour
planes, the isochrone matches the MS stars that have $V$ (or $r$) $\lta 17$ very
well.  As expected, the observed $B-V$ colours of cool, faint stars are more
problematic than their $V-K_S$ colours.  Indeed, the close agreement of the
isochrone with the observations on the $(V-K_S,\,V)$-diagram suggests that the
temperatures of the models are quite good (for additional discussion of this
issue see \citealt{vbf14}).  If this is correct, then the deviations which
are apparent at faint magnitudes in the other CMDs must be telling us something
about the MARCS transformations, or the photometric data, or perhaps that the
assumed metallicity or metals mixture is wrong.  As discussed in the next
section, colours for metal-rich, lower MS stars are especially dependent on
[Fe/H] and [$\alpha$/Fe] (and presumably on any metal with a high abundance).  
(We note, in closing this section, that fits of completely independent
isochrones to the Sloan data for M$\,$67 were obtained by \citealt{apm09}.)

\section{Morphology of the CMD in various colour planes and as function of
reddening}\label{sec:morpho}

\begin{figure*}
\begin{center}
\includegraphics[width=0.9\textwidth]{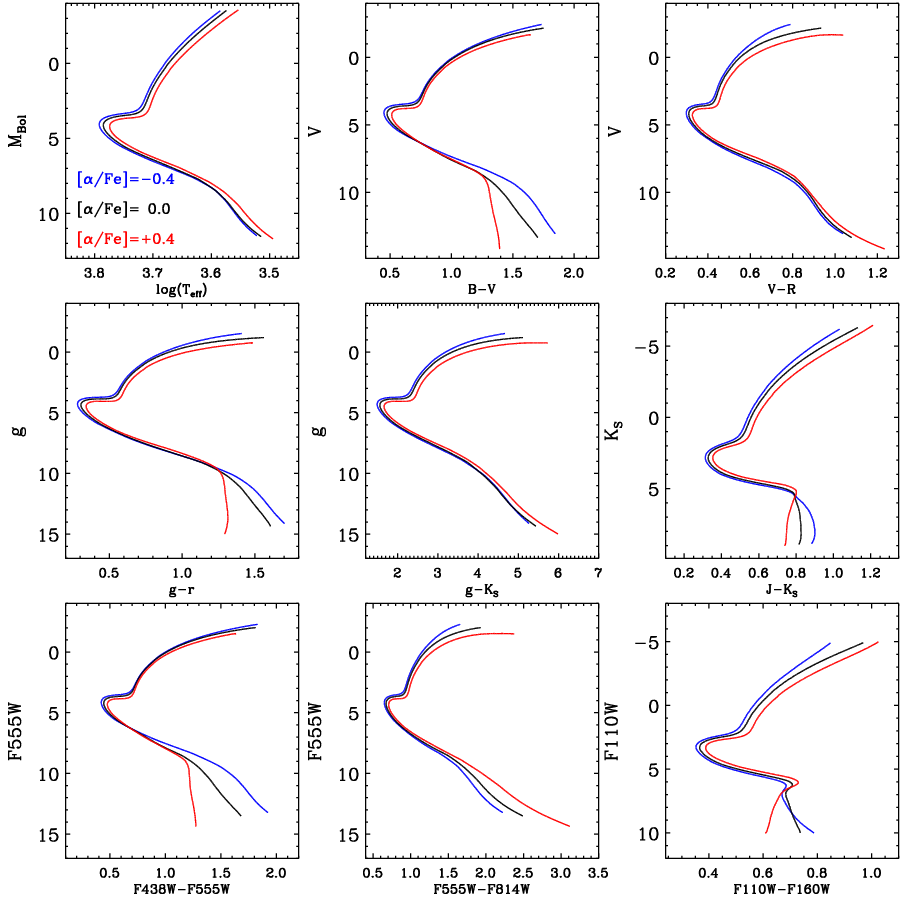}
\caption{Morphology of the CMD in different combinations of filters for a 
$12$~Gyr isochrone having $Y=0.25$, $\feh=-0.8$ and three different values of 
$\aFe=-0.4, 0$ and $+0.4$.}\label{f:cmd_m08}
\end{center}
\end{figure*}

With our set of stellar isochrones and colours, it is of some interest to
study how variations in chemical composition --- $\aFe$, to be specific --- 
affect the morphology of the CMD when using different filter combinations.
For this purpose, we consider two different cases. In Figure \ref{f:cmd_m08},
we examine the effects of changing the $\alpha$-element abundances on the
location and shape of an $\feh=-0.8$, $12$~Gyr isochrone, which is roughly
representative of moderately metal-rich globular clusters in the Milky Way.
The upper left-hand panel shows three isochrones in the theoretical
$\log\teff-M_{\rm Bol}$ diagram, all for the same iron content, but different
values of $\aFe$. Because of this difference, the total mass-fraction abundance
of the metals ($Z$) is somewhat different in the three isochrones, which
explains why they are slightly offset from each other.  However, apart from
this, there are no morphological changes. 

The other panels of Figure \ref{f:cmd_m08} show the same three isochrones, but
as transformed to various CMDs that involve the magnitudes predicted for
various optical and infrared filters.  In nearly all cases, the only part of
the CMD that is substantially affected by variations in the abundances of the
$\alpha$-elements is the lower-MS (at $\teff \lesssim 4000$~K). A
detailed explanation of the cause(s) of this sensitivity is beyond the scope
of this paper, but it stems essentially from the onset of various molecules 
(e.g., $\rm{H_2O}$ and TiO, which form more easily with increasing $\aFe$)
as important sources of blanketing.  Interestingly, recent $F110W$ and $F160W$
photometry of two globular clusters (namely, NGC$\,$2808 and M$\,$4, which both
have $\feh \simeq -1.2$, e.g., \citealt{cbg09}, making them about 0.4 dex more 
metal-poor than the isochrones considered here) has revealed that they have 
split lower main sequences 
\citep{mmc12,mmb14}.  Indeed, the observed morphologies are
qualitatively quite similar to the results shown in the bottom right-hand
panel, which suggests that star-to-star variations in the abundances of one
or more of the $\alpha$-elements is responsible for this phenomenon.  It may
be the case, for instance, that a split main sequence is one of the
consequences of the presence of O-rich and O-poor populations within a given
globular cluster.  Further work is clearly needed to check out this
speculation, but the main point of this example is that chemical abundance
anomalies can be expected to leave detectable photometric signatures.

\begin{figure*}
\begin{center}
\includegraphics[width=0.9\textwidth]{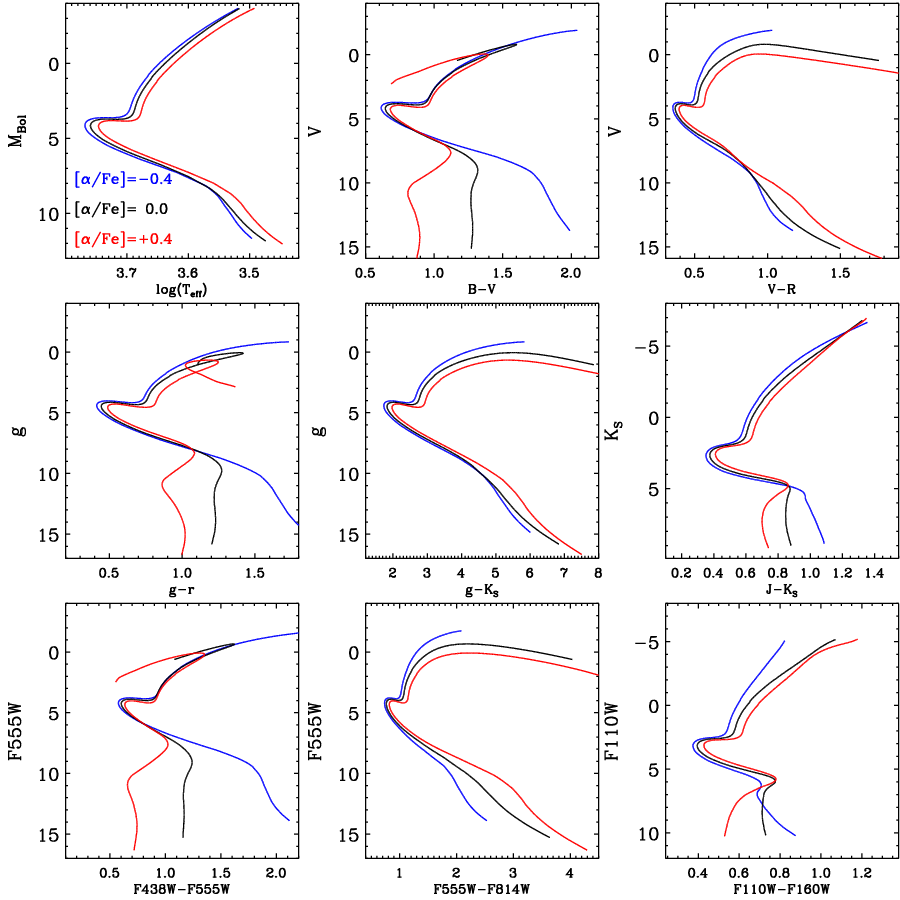}
\caption{Same as previous Figure, but using instead a $10$~Gyr isochrone 
having $Y=0.25$ and $\feh=0.0$.}\label{f:cmd_p00}
\end{center}
\end{figure*}

Figure \ref{f:cmd_p00} is identical to the previous one (in particular, the
same variations of $\aFe$\ are considered), except that somewhat younger
isochrones (10~Gyr) for $\feh=0.0$ have been plotted.  The split in the 
lower-MS is even more dramatic, and certain filter combinations reveal
interesting features for the upper CMD (above the TO) as well.  As the
abundances of the $\alpha$-elements increase, the RGB swings to blue and red 
colours
and even makes loops in the CMD in some cases. Again, a detailed discussion
of the causes of such effects is beyond our purposes, but a visual inspection
of synthetic RGB optical spectra provides a clue as to what is going on.  At
low $\alpha$-element abundances ($\aFe=-0.4$), the overall shapes of spectra
essentially scale with decreasing $\teff$ (implying that, e.g., the $g-r$
index gets monotonically redder).  However, in the $\alpha$-enhanced models,
there is a dramatic dampening of the flux with decreasing $\teff$, likely
driven by the strengthening TiO bandheads.  We suspect that this prevents some
colour indices from getting monotonically redder with decreasing $\teff$. 

While we defer a more detailed explanation of these effects to a subsequent 
investigation, optical spectrophotometry of cool metal-rich giants (e.g., the 
Next Generation Spectral Library --NGSL-- \citealt{ngsl} and the 
Medium Resolution INT Library of Empirical Spectra --MILES-- \citealt{miles})
can be used to assess the 
reliability of synthetic spectra in this regime.
Nevertheless, the apparent high sensitivity of many colours to the temperatures
and chemical abundances of metal-rich stars (both of giants and dwarfs,
depending on the filters used) has considerable potential in advancing our 
understanding of, in particular, the Galactic bulge given that it may be
possible to distinguish between $\alpha$-rich and -poor populations. 

\begin{figure*}
\begin{center}
\includegraphics[width=0.7\textwidth]{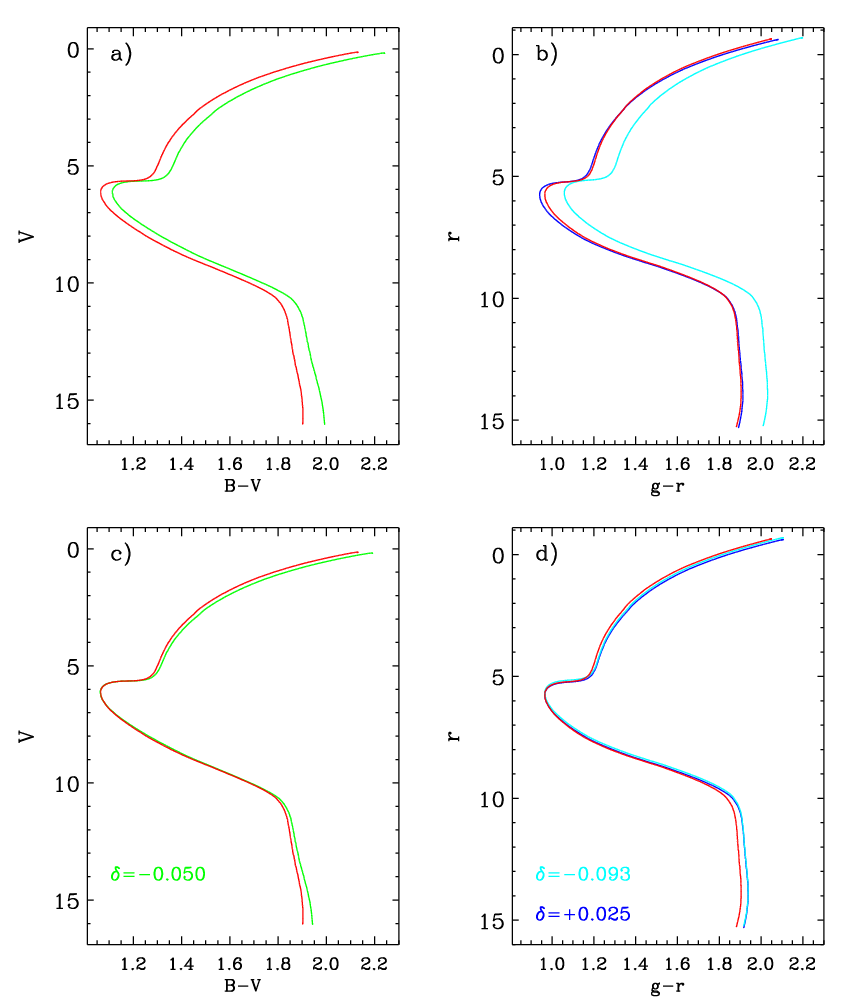}
\caption{Effect of reddening $E(B-V)=0.6$ on a 12~Gyr, $Y=0.25$, $\feh=-0.8$, 
$\aFe=+0.4$ isochrone when using our published colours (in red), or assuming 
constant extinction coefficients (green, blue and cyan). See text for 
discussion.}\label{f:cmd_ebv}
\end{center}
\end{figure*}

Our final topic of discussion concerns the effects of extinction on the CMD.
In most investigations, the extinction coefficient in a given band $R_\zeta$
is assumed to be constant for stars of different spectral types. That is, once
a value of $E(B-V)$ is known, the colour excess in any colour other than $B-V$
is determined from 
$E(\zeta-\eta) = A_\zeta - A_\eta = (R_\zeta - R_\eta) E(B-V)$, from which the 
unreddened colour index is
\begin{equation}\label{eq:unredcol}
(\zeta-\eta)_0 = (\zeta-\eta) - E(\zeta-\eta). 
\end{equation}
Similarly, for magnitudes 
\begin{equation}\label{eq:unredmag}
m_{\zeta,0} = m_\zeta - A_\zeta = m_\zeta - R_\zeta E(B-V), 
\end{equation}
and the distance modulus in a given band $\zeta$ (affected by extinction 
$A_\zeta$) is related to the true distance modulus via
$(m-M)_0 = (m-M)_\zeta - A_\zeta$. 
In Figure \ref{f:cmd_ebv}, we show in red a metal-poor isochrone on different
colour-magnitude planes that has been reddened at each point with an input 
$R_V=3.1$ extinction law, assuming a nominal $E(B-V)=0.6$. As discussed in
Section \ref{sec:ext}, our nominal extinction coefficient and colour excess are 
valid for early type stars.
In panel a) we show in green the same isochrone, but in this case the reddening
has been calculated assuming the same colour excess ($0.6$) and $R_V=3.1$ in 
the relations
(\ref{eq:unredcol}) - (\ref{eq:unredmag}) above.  Likewise, in panel b) we
show the effect of using the constant SDSS extinction coefficients provided by
\citet[][in blue]{apm09} and \citet[][in cyan]{McCall04}. The extinction
coefficients of \cite{apm09} closely match the absolute position of our
reddened isochrone, simply because those coefficients were derived using a
synthetic spectrum for $\logg=4.5$, $\feh=0.0$ and $\teff=5750$~K, which is 
similar to those that are appropriate for our reference isochrone. In contrast,
the extinction coefficients given by \cite{McCall04} were derived using the
$\alpha$\,Lyr reference spectrum, which is responsible for the relatively
large offset between the cyan and red isochrones.  These two panels thus
illustrate the risk involved in trying to fit isochrones to photometric data
on the assumption of a literature value of $E(B-V)$ when a slightly different  
value may actually be more correct for the stars under investigation.

In the lower panels c) and d), we re-register all of the isochrones to match
our reference isochrone at the TO point, applying whatever colour shifts 
$\delta$ are required (as noted). Alternatively, it would be possible to 
obtain the same result using in the relations (\ref{eq:unredcol}) - 
(\ref{eq:unredmag}) the extinction and reddening coefficient appropriate for 
a moderately metal-poor turn-off star (i.e., $E(B-V)=0.55$ and $R_V=3.40$ for 
the case of panel c) and $E(B-V)=0.55$, $E(g-r)=0.62$ and $R_r=2.95$ for the 
case of panel d), see Section \ref{sec:ext} and Appendix \ref{app}). 
Nevertheless, it is clear from these panels that, despite
this renormalization, the morphologies of the CMDs for the different cases 
are not identical along the lower-MS or the RGB.
While these subtleties are usually negligible in the presence of low reddening, 
this is not the case for high values of the extinction.  We emphasize that our
reddened colours account for the effects of extinction in a fully
self-consistent way.  We also reiterate our comment that, when using literature
values of $E(B-V)$, attention should be paid to how they were derived in order
to minimize the inconsistencies highlighted above.  More often than not, small
adjustments to literature values of $E(B-V)$ might be applied. However, this 
should not be regarded as an excuse to use reddening as a free parameter, 
since an error in the reddening is only one of many possible reasons why an 
isochrone may not reproduce the observed TO colour of a cluster CMD even when
well-supported estimates of the distance, metallicity, and $E(B-V)$ are 
adopted.

\section{Conclusions}

The large grids of MARCS model atmospheres, now available for 
different chemical compositions, together with the fluxes calculated at 
$\sim 10^5$ wavelength points, represent an especially valuable resource for 
the study of stars and stellar populations \citep{g08}. 
To facilitate their application to photometric data, we have provided the 
means to generate, and to interpolate in, tables of synthetic BCs (and thus 
colours) applicable to several broad-band systems, for a wide range of 
physical and chemical abundance parameters (see Appendix \ref{app} for detailed 
information on the parameter space covered). The photometric systems for which 
such transformations are provided 
include Johnson-Cousins-2MASS ($UBVR_CI_CJHK_S$), Sloan ($ugriz$), and 
HST-ACS/WFC3 (26 different filters, each in the {\tt ST}, {\tt AB}, and 
{\tt VEGA} mag systems). The differences between the latter have been fully 
described, and the most updated photometric zero-points and absolute 
calibrations used. A novel feature of our transformations is that 
they can be corrected for the effects of reddening (for any $E(B-V)$ value 
between $0.0$ and $0.72$, inclusive) in a fully consistent way;
i.e., the variation in the colour excess with the spectral type of a star is
correctly taken into account. So far, our results assume the 
standard $R_V=3.1$ for the extinction law, although the inclusion of other 
values (as e.g.\,more appropriate towards certain Galactic sightlines) is 
currently underway. The impact of varying the assumed microturbulent
velocity on synthetic magnitudes in UV, optical, and IR bandpasses has also
been examined in some detail: all of our results assume $\xi = 2$~km/s. While 
this value is generally appropriate for stars across most of the CMD, its 
variation can seriously impact colours for $\teff \lesssim 4000$~K, and 
increasing metallicities, in particular towards the UV. 

Encouragingly, we have found that the MARCS colour--$\teff$--metallicity 
relations and BCs in most of the broad-band filters agree with those derived 
empirically above $\simeq 4500$~K using the IRFM \citep{c10} to within a 
couple/few percent. This is easily within the uncertainties associated with 
the zero-points of the photometric systems and of the temperature scale.
Moreover, the predicted metallicity dependence of such relations appears to
reproduce observed trends quite well according to our analyses of a large
sample of solar neighbourhood stars that have $ugriz$ photometry. In general, 
the fits of isochrones to globular and open cluster CMDs (specifically, those 
of M$\,$5, NGC$\,$6791, and M$\,$67) yielded consistent interpretations of the 
data on different colour planes, though small (0.01-0.02 mag) colour offsets 
were sometimes found in some CMDs relative to others for the same cluster. 
Even the synthetic $U$ magnitudes did not appear to be seriously discrepant, 
at least for metal-poor TO stars. Larger deficiencies might affect synthetic 
colours of late M-stars, although this regime has only been partly explored 
in this work (and certainly not at the coolest $\teff$ of the MARCS grid).

Our brief consideration of Victoria-Regina isochrones on many different CMDs 
revealed
the strong sensitivity of the colours of lower-MS stars to
[$\alpha$/Fe], which undoubtedly has relevance for the bifurcated CMDs of
M$\,$4 and NGC$\,$2808. Not surprisingly, the colours of cool stars, are very 
dependent on the chemical composition. As such, changing $\aFe$ can modify 
quite dramatically the morphology of the CMD at its coolest extremities, both 
along the lower-MS and upper-RGB.

We defer to a future investigation a more detailed exploration of the effects 
of chemical abundances on colours, in particular accounting for the 
anti-correlations observed in many globular clusters \citep[e.g.,][]{ssw11}.
In this respect, it will also be interesting to extend our grid of synthetic 
BCs to intermediate and narrow-band systems, where synthetic models seem to 
perform less satisfactorily in comparison with observations 
\citep[e.g.,][]{onehag09,melendez10}. The Str\"omgren system will also be of 
particular interest, because of its capability to help distinguish between
multiple populations in globular clusters \citep[e.g.,][]{gva98,yong08,cbg11}.

\section*{Acknowledgements}
We thank the referee for a careful reading of the manuscript and constructive 
comments. Useful discussions with Mike Bessell and Remo Collet, and guidance 
from Peter Bergbusch in creating figures A2 and A3 in Appendix A are 
acknowledged with gratitude. The contributions of DAV to this project were
supported by a Discovery Grant from the Natural Sciences and Research Council
of Canada.  This research used the facilities of the Canadian Astronomy Data
Centre operated by the National Research Council of Canada with the support of
the Canadian Space Agency.

\bibliographystyle{mn2e}
\bibliography{refs}

\appendix

\section[]{Interpolation routines}\label{app}

\begin{figure}
\begin{center}
\includegraphics[width=0.48\textwidth]{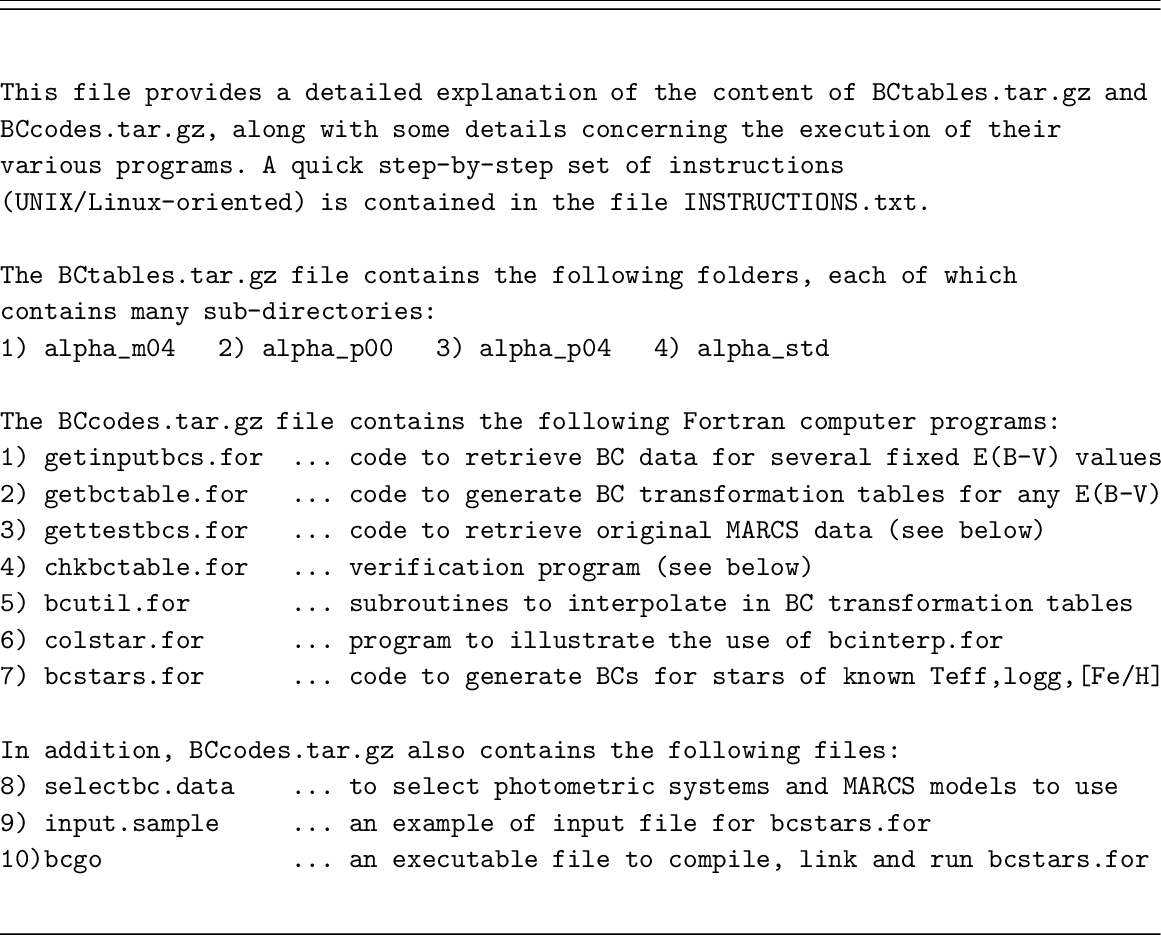}
\caption{Excerpt of the {\tt README} file provided with our package.}
\label{fig:A0}
\end{center}
\end{figure}

The transformations described in this paper and the codes (in FORTRAN) that 
are needed to tabulate, and interpolate in the BCs for the filters and 
reddenings of interest are contained in the two files {\tt BCtables.tar.gz} 
and {\tt BCcodes.tar.gz}, repectively. A detailed explanation of their content 
(and use) is provided in a {\tt README} file, an excerpt of which is shown in 
Figure \ref{fig:A0}. All these files may be obtained from CDS.

Before describing their contents, a few general remarks are in order.
Although the MARCS model atmospheres and synthetic spectra were computed for
$-5.0 \le$ [Fe/H] $\le +1.0$, $-0.5 \le \logg \le 5.5$, $2500 \le \teff \le
8000$ K, and $-0.4 \le$ [$\alpha$/Fe] $\le +0.4$, the coverage of parameter
space is not complete.  For instance, there are no model atmospheres at low
gravities and high temperatures (which would be of limited usefulness, anyway),
and relatively few computations at the lowest temperatures, gravities, and
metallicities, or at $\logg = 5.5$.  Moreover, the number of $\teff$ values
for which model atmospheres were generated varies with $\logg$, [Fe/H], and
[$\alpha$/Fe].  To deal with this nonuniformity, which obviously presents some
challenges for the setting up of interpolation tables, cubic splines were used
whenever possible to determine the transformations, which consist of BCs for 
each filter bandpass that we have considered, when model
atmospheres were lacking at gravities and temperatures in the middle of their
respective ranges. Linear or 3-point extrapolations were also made, as needed,
to ``fill in" missing BC entries located near the upper and lower limits of the
dependent variables.

\begin{figure}
\begin{center}
\includegraphics[width=0.48\textwidth]{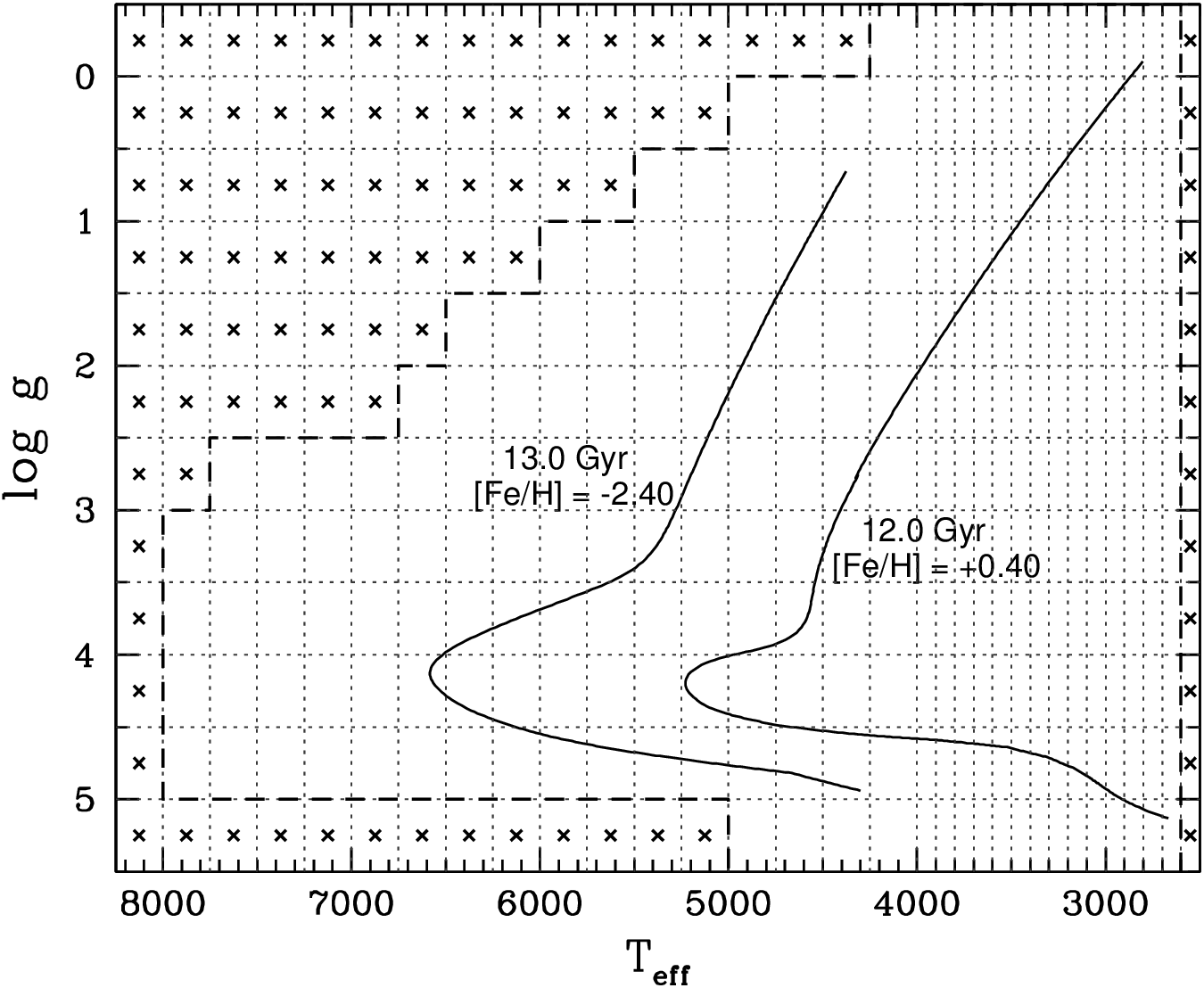}
\caption{BCs are given for $\teff$\ and $\logg$\ values
within the region enclosed by the dashed lines.  (Tables of the same dimensions
are provided for each of the grid values of [Fe/H] and [$\alpha$/Fe], which
vary from $-4.0$ to $+1.0$ and $-0.4$ to $+0.4$ (for a subset of the [Fe/H]
values), respectively (see the text).  Isochrones (from \citealt{vbf14}) have
been plotted to show that the transformations provide sufficient temperature and
gravity coverage for stellar models over a wide range in mass ($\gta 0.12
\msol$) and evolutionary state.}
\label{fig:fig21}
\end{center}
\end{figure}

In order to minimize the number of extrapolations, while maximizing the size of
the transformation tables, the decision was made to restrict the latter to the
ranges in gravity and temperature shown in Figure~\ref{fig:fig21} (the region
enclosed by the dashed lines).  Victoria-Regina isochrones (\citealt{vbf14}) for
[Fe/H] $= -2.4$ and $+0.4$, which include masses as low as $0.12 \msol$ and
extend to the RGB tip, have been included in this plot simply to illustrate 
where such models are located in this diagram.  In this figure, the grid values
of $\logg$ and $\teff$\ are indicated by the intersection of the dotted grey
lines, and it is for these values that BCs are provided.  (Tables of exactly
the same dimensions have been created for all of the grid values of [Fe/H] and
[$\alpha$/Fe].)  Thus, no data are provided for $\teff < 2600$\ K,
while BCs for $\logg = 5.5$ are given only for $5000 \le \teff \le 2600$\ K,
etc.  Perhaps the most important limitation of the transformation tables is
that they can be used only for $\teff \le 8000$\ K, which is a consequence of
the fact that MARCS model atmospheres were not computed for higher temperatures.
As a result, they cannot be used for, e.g., hot MS or blue horizontal-branch
stars --- but hopefully an extension of the tables to higher temperatures will
be made at some future date.  

\begin{figure}
\begin{center}
\includegraphics[width=0.48\textwidth]{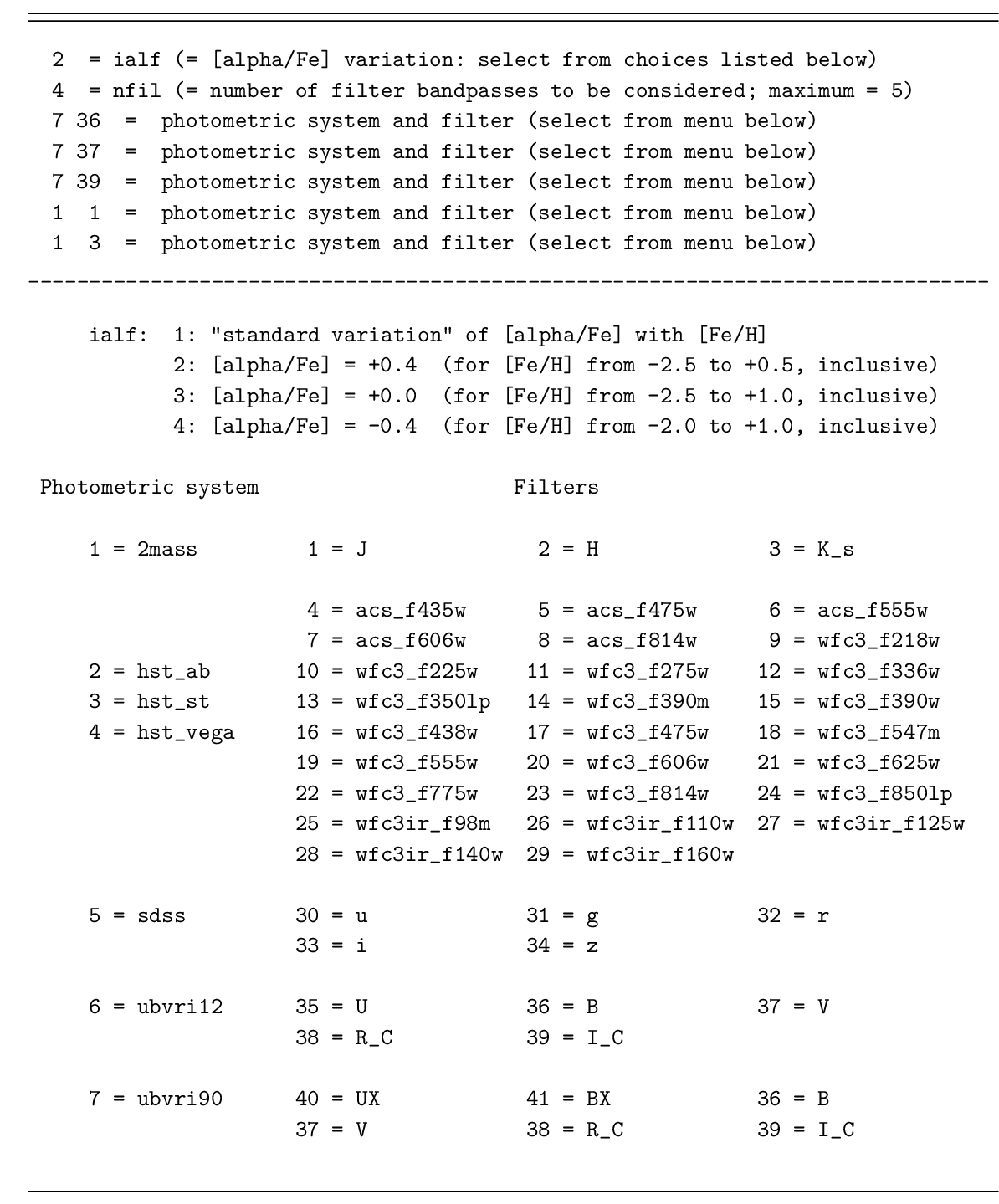}
\caption{A sample {\tt selectbc.data} file: it is used to choose among four
possible variations of [$\alpha$/Fe] with [Fe/H] (via the parameter {\tt ialf}),
to select the number of filters for which BC transformations are sought ({\tt
nfil}), and to identify the particular photometric system(s) and filter(s) of
interest (using lines 4--7).  The possible options are listed below the
horizontal dashed line.}
\label{fig:fig22}
\end{center}
\end{figure}

Once {\tt BCtables.tar.gz} has been retrieved from CDS, it would be
prudent to open it in a separate directory as it contains a fairly extensive set
of sub-directories (each containing the BCs for a specific photometric system).
{\tt BCcodes.tar.gz}, which contains the software that manipulates the
transformation data, should be opened in the same "home" directory because the
computer programs assume a particular directory structure for the data files
and they will not execute properly unless that structure is in place.  Their
execution is controlled by the included {\tt selectbc.data} file, a sample
of which is shown in Figure~\ref{fig:fig22}.  It essentially provides a menu
(the information given below the dashed line) that enables the user to select up
to 5 filters for which a table of BCs will be created for subsequent
interpolations (so as to, e.g., transform stellar models from the theoretical
plane to various CMDs of interest).  The main advantage of tabulating BCs,
instead of colours, is that the former is a much more compact way of presenting
the transformations.  For instance, from 5 BCs for the Johnson-Cousins
$UBV(RI)_C$ system, 10 different colour indices ($U-B$, $U-V$, $\ldots$ $B-V$,
$\ldots$ $R_C-I_C$) can be derived.  (Synthetic absolute magnitudes can be 
calculated from $M_\zeta = M_{\rm bol} - BC_\zeta$ for the $\zeta^{th}$ filter, 
and colours may be evaluated from either $\zeta - \eta = M_\zeta - M_\eta$ or 
$\zeta - \eta = BC_\eta - BC_\zeta$, where $\zeta$ represents a bluer bandpass 
than $\eta$.)

The integer on the first line of {\tt selectbc.data} chooses among the four
possible variations of [$\alpha$/Fe] with [Fe/H] for which BC transformations
are available.  If {\tt ialf} $= 1$, the so-called ``standard" variation is
selected: this assumes [$\alpha$/Fe] = $+0.4$ for $-4.0 \le$ [Fe/H]
$\le -1.0$, [$\alpha$/Fe] $= 0.0$ for $0.0 \le$ [Fe/H] $\le +1.0$, and
[$\alpha$/Fe] $= -0.4$ for $-1.0 <$ [Fe/H] $< 0.0$. (Note that BCs are
not provided for [Fe/H] $< -4.0$, even though some MARCS model atmospheres were
computed for [Fe/H] $= -5.0$.  There were simply too few of the latter cases
to cover the adopted ranges in $\logg$ and $\teff$\ without having to carry
out excessive numbers of extrapolations.)  Setting {\tt ialf} $= 2, 3,$ or 4
will result in the retrieval of the BCs for, in turn, [$\alpha$/Fe] $= +0.4$
for $-2.5 \le$ [Fe/H] $\le +0.5$, [$\alpha$/Fe] $= 0.0$ for $-2.50 \le$ [Fe/H]
$\le +1.0$, or [$\alpha$/Fe] $= -0.4$ for $-2.0 \le$ [Fe/H] $\le +1.0$. 

The second line of {\tt selectbc.data} defines {\tt nfil}, which can have a
maximum value of 5.  In many applications, the magnitude on the $y$-axis of a
CMD is also used to compute the colour on the $x$-axis, as in the case of a
$(B-V,\,V)$-diagram; in which case, {\tt ialf} would be set to a value of 2.
However, {\tt nfil} $= 5$ would be the natural choice for observations obtained 
in, e.g., the $ugriz$ system.  The next 4--7 lines specify the photometric
system(s) and filters of choice.  In the example given in Fig.~\ref{fig:fig22},
the 1990 calibration (named {\tt ubvri90}, see Section \ref{subsec:jc2m}) 
of the $B$, $V$, and $I_C$ filters,
along with the 2MASS $K_S$ filter, have been selected.  (The $7^{th}$ line,
which specifies the integers 1 and 3, will not be read in this case because
{\tt nfil} $= 4$.)  The photometric systems and filters can be identified
in any order.

Once the parameters in {\tt selectbc.data} have been set, two programs must
be executed in order to obtain the desired BC transformation tables.  The
first one that needs to be compiled and executed is {\tt getinputbcs.for}.
Because reddening-corrected BC values have been derived for $E(B-V) = 0.00$
to 0.72, in steps of 0.02, the user can request transformations for any
reddening within this range.  However, rather than storing such a large
dataset, only the tables for $E(B-V) = 0.00, 0.12, 0.24, \ldots 0.72$
have been retained.  We found that spline interpolations in tables for these
7 reddenings were able to reproduce the reddening-adjusted transformations
in the full set of tables to within 0.001 mag, independently of the selected
filter.  Consequently, the purpose of {\tt getinputbcs.for} is to retrieve
from the appropriate sub-directories the BCs for $0.0 \le E(B-V) \le 0.72$, 
in steps of 0.12 mag.  These data are placed in the seven files that have
been given the names \verb+inputbc_r00.data+, \verb+inputbc_r12.data+, etc.
(These files are rewritten each time {\tt getinputbcs} is executed.)

The next step is to compile and execute {\tt getbctable.for}, which prompts
the user for a value of $E(B-V)$ and then interpolates in the
\verb+inputbc_*.data+ files to produce the requested BC transformations for
that choice of the reddening.  Since the ranges in [Fe/H] are different for the
cases that are specified by the parameter {\tt ialf} (see above), the output
file generated by {\tt getbctable} is given the name \verb+bc_std.data+ if
{\tt ialf} $= 1$, \verb+bc_p04.data+ if {\tt ialf} $= 2$, \verb+bc_p00.data+
if {\tt ialf} $= 3$, or \verb+bc_m04.data+ if {\tt ialf} $= 4$.  Perhaps the
main reason for doing this is to facilitate interpolations in the resultant
transformations for arbitrary values of [$\alpha$/Fe] within the range $-0.4
\le$ [$\alpha$/Fe] $\le +0.4$.  (To perform such interpolations, it is clearly
necessary to cycle through the above procedure in order to create the respective
\verb+bc_*.data+ files corresponding to {\tt ialf} $= 2$, 3, and 4.)  Note that
{\tt getinputbcs} and {\tt getbctable} are completely self-contained; i.e., all
file assignments are done internally.  In summary, to obtain reddening-adjusted
BC transformations for up to 5 different filters, one simply sets the parameters
in {\tt selectbc.data} to the desired values, and then executes
{\tt getinputbcs} and {\tt getbctable}.

\begin{figure}
\begin{center}
\includegraphics[width=0.48\textwidth]{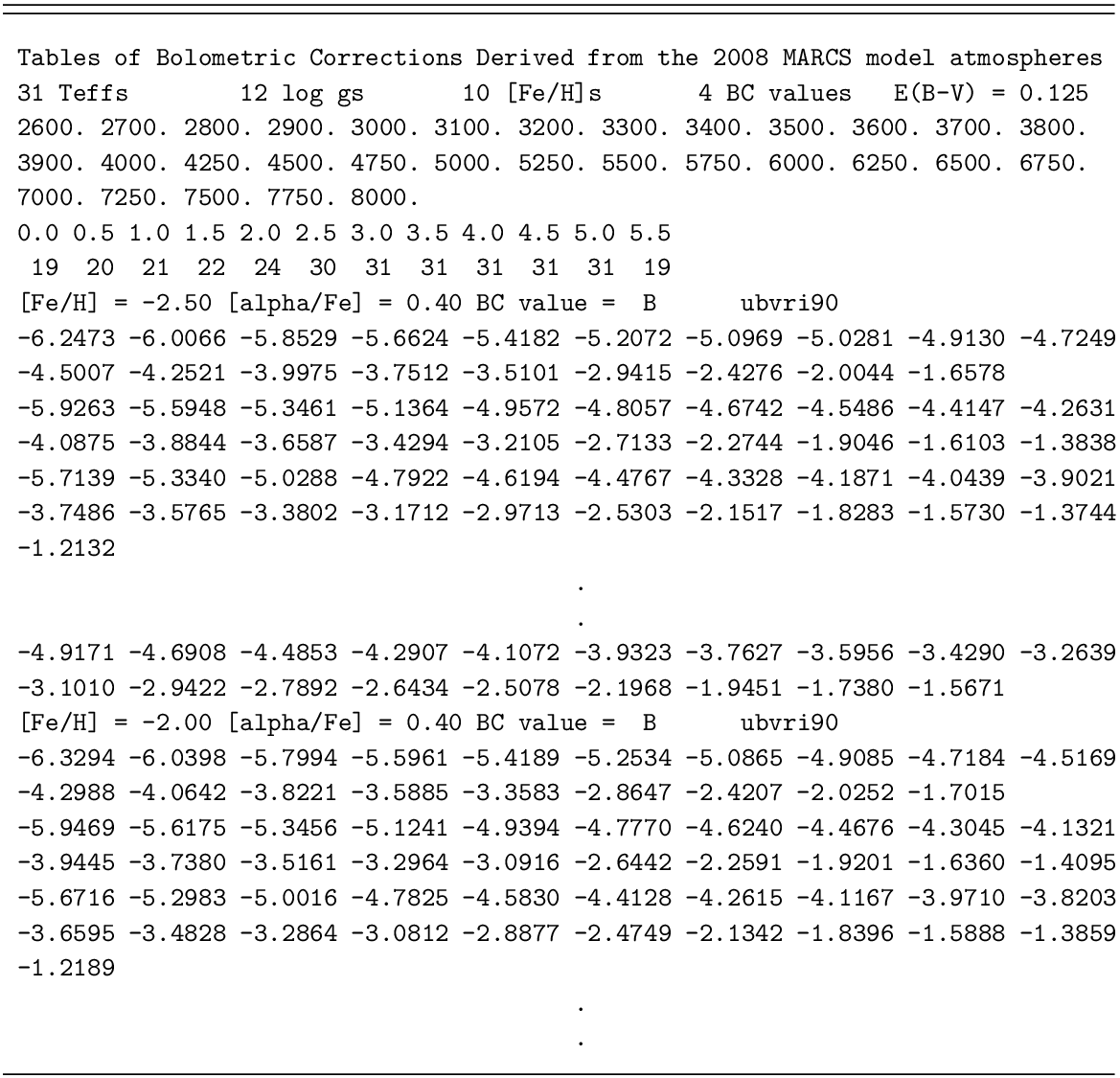}
\caption{The first $\sim 25$\ lines of the output file that is produced if
{\tt getinputbcs} and {\tt getbctable} are executed when the values of the
initializing parameters in {\tt selectbc.data} are set to the values shown
in the previous figure. As indicated in the second line, these transformations 
assume $E(B-V)=0.125$, which is the reddening value that was entered in 
response to a prompt that is issued by {\tt getbctable}.}
\label{fig:fig23}
\end{center}
\end{figure}

The output data files so obtained have been organized in a straightforward way.
Figure~\ref{fig:fig23} reproduces the beginning of the \verb+bc_p04.data+ file
that would be created using the example of {\tt selectbc.data} given in the
previous figure.  The first (title) line is included primarily to reinforce
the fact that the quantities in the tables are BCs (not colours).  The second
line specifies the number of $\teff$s, gravities, [Fe/H] values, and BCs, as
well as the reddening value that was entered when prompted for it by
{\tt getbctable}.  The next four lines list the grid values of $\teff$ and
$\logg$, while the following ($7^{th}$) line specifies how many BC values,
beginning with the first entry at 2600 K, are included in the table at each
$\logg$ value.  For example, the minimum gravity in this table is $\logg
= 0.0$, and according to the first entry on the seventh line, BC transformations
are provided for the first 19 values of $\teff$\ (i.e., up to 5000 K), which is
consistent with what is shown in Fig.~\ref{fig:fig21}.  The seventh line thus
helps to describe the irregular shape of the region in $\teff$, $\logg$ space
where interpolations are permitted. 

The rest of the sample \verb+bc_p04.data+ file lists, in the form of
mini-tables, BCs in the order of increasing [Fe/H], and within each table, they
are organized in rows for the grid values of $\logg$, with a maximum of $10$
numbers per row.  Hence the 19 BC values at $\logg = 0.0$, in the order of
increasing $\teff$, are contained in the two rows immediately after the header
line that gives the values of [Fe/H] and [$\alpha$/Fe] which are applicable to
the BC data that follow.  All of the transformations for a given filter, which
is also identified in the mini-table header lines are grouped together.  Thus,
all of the transformations for the {\tt ubvri90} $B$ filter are listed before
all of those for the $V$ filter (which are likewise organized in the direction
of increasing [Fe/H]), etc.

In addition to subroutines (such as \verb+getbcs_p04.for+) that perform
interpolations in the BC tables, {\tt BCcodes.tar.gz} contains programs that
can be used to verify that everything is working correctly.  To be more
specific, {\tt gettestbcs.for} retrieves from the appropriate sub-directories,
the original MARCS transformations that were used in the construction of the 
interpolation tables (for one of six possible values of $E(B-V)$; namely, $0.0$,
$0.08$, $0.14$, $0.28$, $0.44$, or $0.56$). These data are stored in files 
with names
like \verb+sph00.data+ or \verb+ppl40.data+, where the prefixes {\tt sph} and
{\tt ppl} indicate that the data are based on ``spherical" or ``plane-parallel"
model atmospheres, respectively, and the two digits indicate the $\logg$ value
(e.g., {\tt 00} implies 0.0 whereas {\tt 40} implies $4.0$; $\logg = -0.5$ is
coded as {\tt m05}). These values of $E(B-V)$ were chosen so as to be offset
by different amounts from the values ($0.0$, $0.12$, $0.24$, $0.36$, $\ldots$) 
used in the spline fits.

The second of three steps in the verification process is to run 
{\tt getbctable} for the same value of the reddening.  In the final step, 
which requires the execution of
{\tt chkbctable.for}, the [Fe/H], $\teff$, and $\log\,$g values listed in the
\verb+sph*.data+ and \verb+ppl*.data+ files are used to obtain BC values, by
interpolating in the output file from {\tt getbctable}: these are then compared
with the values in the original MARCS tables (i.e., \verb+sph*.data+, etc.).
In our experience, the differences between the interpolated and tabulated BCs
are always $< 0.001$ mag: they tend to be the largest for blue and UV 
bandpasses. Differences will be exactly $0.0$ only when the BCs for 
$E(B-V) = 0.0$ are selected by {\tt gettestbcs} given that exactly the same 
data are employed in the construction of the splines.
A short program, with results, is also included in the package to
illustrate how to perform interpolations in the tables produced by
{\tt getbctable}.  As already pointed out, a detailed explanation of these 
auxiliary codes is contained in a {\tt README} file that is also provided.

Finally, it is worth mentioning that the extinction coefficients
(and reddening values) that apply to reddened stars of different spectral
types may be easily calculated using our interpolation routines. Consider,
for instance, an old, moderately metal-poor ($\feh=-0.8$, $\aFe=+0.4$),
turnoff star ($\teff=5950$~K, $\logg=4.2$) that is affected by a nominal
reddening $E(B-V)=0.6$.  On the assumption of the canonical extinction law,
$R_V=3.1$, the following BCs are obtained in a number of selected bands:
$BC_B=-3.03$, $BC_V=-1.97$, $BC_g=-2.54$, $BC_r=-1.58$.  If that star were
unreddened, the values of the same BCs would be: $BC_B=-0.61$, $BC_V=-0.10$,
$BC_g=-0.30$, and $BC_r=0.04$. It thus follows that, at the turnoff,
$E(B-V)=(-1.97+3.03)-(-0.10+0.61)=0.55$ and $E(g-r)=(-1.58+2.54)-(0.04+0.30)
=0.62$, with extinction coefficients $R_B=\frac{3.03-0.61}{0.55}=4.40$,
$R_V=\frac{1.97-0.10}{0.55}=3.40$, $R_g=\frac{2.54-0.30}{0.55}=4.07$, and
$R_r=\frac{1.58+0.04}{0.55}=2.95$.  Note, in particular, that the difference
$R_B-R_V$ is exactly 1.0 {\it because} the extinction coefficients were
calculated assuming the value of $E(B-V)$ that applies to the star in question.

This difference would be $< 1$ if the $E(B-V$) value appropriate to early-type
stars is adopted --- which is the usual convention for reddening determinations
that are reported in the literature.  Returning to the example considered in 
the previous paragraph, $R_B = \frac{3.03-0.61}{0.6} = 4.03$ and $R_V =
\frac{1.97-0.10}{0.6} = 3.12$, with the result that $R_B-R_V = 0.91$ and
$E(B-V) = (0.91)(0.6) = 0.55$. Reinforcing the point already made in Section
\ref{sec:morpho} \citep[also see][]{b98}, these calculations show that
the $E(B-V)$ value produced by a given amount of dust will be significantly
less for a $5950$~K (late F-type) turnoff star than for early O- and B-type 
stars. Similarly, colour excess ratios will scale according to spectral type 
since, for filters $\zeta$ and $\eta$, the difference $R_\zeta-R_\eta$ is equal 
to $E(\zeta-\eta)/E(B-V)$, where the denominator of this term is the nominal 
reddening.  Clearly, to correct colours (Eq.~\ref{eq:unredcol}) and
magnitudes (Eq.~\ref{eq:unredmag}) for the effects of interstellar extinction,
one should use {\it either} $0.55$ and $(3.4)(0.55)$ {\it or} $(0.60)(0.91)$
and $(3.12)(0.60)$, respectively, in the case of the present example.

In comparisons of isochrones with observed CMDs, it is traditional to apply
constant offsets to the observed magnitudes and/or colours (depending on 
whether an apparent or true distance modulus is assumed) to account for the
effects of foreground gas and dust.  This is much simpler than correcting the
observations (or, more likely, the isochrones) on a point-by-point basis prior
to fitting the models to the data, and indeed, there is no reason to abandon
this practice if, e.g., the main goal of the work is to derive the turnoff
age(s).  (Recall the discussion in Section~\ref{sec:morpho} concerning
Fig.~\ref{f:cmd_ebv}, which showed that, even when the reddening is high, the
two approaches lead to no more than minor differences along the RGB and the
lower MS.)  Since published $R_\zeta$ values are generally based on hot stars
(e.g., Vega, in the case of McCall 2004), we have calculated the extinction
coefficients relevant to turnoff stars ($\logg = 4.1$), $5250 \le \teff \le
7000$~K (in 250~K increments), and $-2.0 \le \feh \le +0.25$ (at the specific
values adopted in the MARCS library).  (This involves nothing more than 
evaluating $A_\zeta$ from the difference in the BCs assuming two nominal 
reddening values, which we choose to be $E(B-V) = 0.0$ on the one hand, and 
$E(B-V) = 0.10$, on the other, and then dividing by $0.10$.)
Variations in $\aFe$ were also considered, but the dependence of
$R_\zeta$ on this abundance parameter was found to be sufficiently weak that
it could be ignored: the results for $\aFe$ $= -0.4, 0.0,$ and $+0.4$
were simply taken to be alternative determinations for the associated [Fe/H]
value.  

Table~\ref{tab:tabler} lists, for the majority of the filters considered in this
study, the mean extinction coefficients so obtained $<R_\zeta>$, along with a 
least-squares fit to the data assuming a functional dependence\footnote{Users 
are warned not to extrapolate these relations outside of the range of their
applicability.  Such caution is advisable even though the dependence of the
extinction coefficients on $\teff$ is usually less dramatic above $10,000$~K
--- as we have found using ATLAS9 models.} 
$R_\zeta=a_0+T_4(a_1+a_2T_4+a_3\feh\,$, where $T_4 = 10^{-4}\teff$. 
With relatively few exceptions, $<R_\zeta>$ reproduces the values of the
extinction coefficients that were computed for the aforementioned values of
$\logg, \teff, \feh$, and $\aFe$ to within 1\%.  The largest
residuals, amounting to $\sim 1.6$\%, were found for the $B$, $g$, and
approximately equivalent {\it HST} filters. Both the maximum residuals and
the {\it rms} deviations, which are typically one-third as large, are reduced
by about a factor of 5 if the least-squares fit to $R_\zeta$ is adopted
instead of the mean value. Only in the case of the $F218W$, $F225W$, $F275W$,
and $F350LP$ filters it was not possible to obtain a good fit to the actual
values of $R_\zeta$; consequently, they have been omitted from this table.
(The mean values were likewise very poor approximations, undoubtedly due to
the strong $\teff$ dependence of the BCs in these four filters.)  In any case,
Table~\ref{tab:tabler} provides two ways of determining approximate (but still
quite accurate) values of $R_\zeta$ that can be used to evaluate the
colour excess ratio for any of the filters that are listed.  Importantly, such
ratios involve the {\it nominal} reddening given that all of our $R_\zeta$
results are based on the standard extinction law, $R_V = 3.1$.

\begin{table*}
\centering
\caption{Values of $R_\zeta$ relevant to turnoff stars with 
$\teff \lesssim 7000\,{\rm{K}}^{a}$. Notice that for a {\it nominal} $E(B-V)$, 
the excess in any given $\zeta-\eta$ colour is 
$E(\zeta-\eta)=(R_\zeta-R_\eta)E(B-V)$, and the attenuation for a 
magnitude $m_\zeta$ is $R_\zeta E(B-V)$. See also discussion in the text.}\label{tab:tabler}
\smallskip
\begin{tabular}{lcccccclccccc}
\hline
\hline 
\noalign{\smallskip}
  & & \multicolumn{4}{c}{$R_\zeta = a_0 + T_4\,(a_1 + a_2\,T_4) + a_3\,\feh$} & & & & \multicolumn{4}{c}{$R_\zeta = a_0 + T_4\,(a_1 + a_2\,T_4) + a_3\,\feh$}\\
Filter & $< R_\zeta >$ & \multispan4\hrulefill &  \phantom{~~~~~~~~~~~~~~~~~~~~~} & Filter & $< R_\zeta >$ & \multispan4\hrulefill \\
 & & $a_0$ & $a_1$ & $a_2$ & $a_3$ & &  & & $a_0$ & $a_1$ & $a_2$ & $a_3$ \\
\noalign{\smallskip}
\hline
\noalign{\smallskip}
\noalign{\centerline{\tt 2MASS}} & & & \\
 \phantom{F4}$J$ & 0.899 & 0.9095 & $-0.0526$    & \phantom{+}$0.0583$ & --- & &
 \phantom{F4}$H$ & 0.567 & 0.5611 & \phantom{+}$0.0176$ & $-0.0115$    & --- \\
 \phantom{F4}$K_S$ & 0.366 & 0.3667 & $-0.0026$    & \phantom{+}$0.0021$ & --- & &
             &       &        &              &              &     \\
\noalign{\vskip -0.1in} \\
\noalign{\centerline{\tt ACS}} \\
 $F435W$ & 4.106 & 3.4929 & \phantom{+}$1.7330$ & $-1.2071$ & $-0.0135$  & &
 $F475W$ & 3.720 & 3.1735 & \phantom{+}$1.4136$ & $-0.8569$ & $-0.0078$  \\
 $F555W$ & 3.217 & 3.0197 & \phantom{+}$0.4900$ & $-0.2738$ & $-0.0018$  & &
 $F606W$ & 2.876 & 2.5200 & \phantom{+}$0.8308$ & $-0.4074$ & $-0.0018$  \\
 $F814W$ & 1.884 & 1.7763 & \phantom{+}$0.2522$ & $-0.1227$ &    ---     & &
         &       &        &              &           &            \\
\noalign{\vskip -0.1in} \\
\noalign{\centerline{\tt WFC3$^{b}$}} \\
 $F336W$  & 5.148 & 4.9002 & \phantom{+}$0.6329$ & $-0.3806$ & $-0.0062$ & &
 $F390M$  & 4.503 & 4.3391 & \phantom{+}$0.5133$ & $-0.4007$ & $-0.0015$ \\
 $F390W$  & 4.432 & 3.6722 & \phantom{+}$2.3798$ & $-1.8697$ & $-0.0165$ & &
 $F438W$  & 4.135 & 3.8431 & \phantom{+}$0.8230$ & $-0.5695$ & $-0.0055$ \\
 $F475W$  & 3.698 & 3.1588 & \phantom{+}$1.4014$ & $-0.8529$ & $-0.0071$ & &
 $F547M$  & 3.140 & 3.0439 & \phantom{+}$0.2472$ & $-0.1483$ & $-0.0015$ \\
 $F555W$  & 3.274 & 2.8862 & \phantom{+}$0.9399$ & $-0.5000$ & $-0.0023$ & &
 $F606W$  & 2.894 & 2.5345 & \phantom{+}$0.8514$ & $-0.4306$ & $-0.0020$ \\
 $F625W$  & 2.678 & 2.5954 & \phantom{+}$0.1656$ & $-0.0481$ &    ---    & &
 $F775W$  & 2.049 & 2.0174 & \phantom{+}$0.0527$ & $-0.0003$ &    ---    \\
 $F814W$  & 1.897 & 1.8027 & \phantom{+}$0.2058$ & $-0.0818$ &    ---    & &
 $F850LP$ & 1.457 & 1.3997 & \phantom{+}$0.1673$ & $-0.1192$ &    ---    \\
 $F098M$  & 1.290 & 1.2709 & \phantom{+}$0.0432$ & $-0.0193$ &    ---    & &
 $F110W$  & 1.047 & 0.9272 & \phantom{+}$0.2848$ & $-0.1428$ &    ---    \\
 $F125W$  & 0.891 & 0.8675 & \phantom{+}$0.0394$ & $-0.0027$ &    ---    & & 
 $F140W$  & 0.753 & 0.7180 & \phantom{+}$0.0754$ & $-0.0290$ &    ---    \\
 $F160W$  & 0.634 & 0.6196 & \phantom{+}$0.0301$ & $-0.0100$ &    ---    & &
          &       &        &              &           &           \\
\noalign{\vskip -0.1in} \\
\noalign{\centerline{\tt SDSS}} \\
 \phantom{F4}$u$ & 4.862 & 4.5725 & \phantom{+}$0.9248$ & $-0.7397$ & $-0.0051$ & &
 \phantom{F4}$g$ & 3.762 & 3.2679 & \phantom{+}$1.2843$ & $-0.7840$ & $-0.0076$ \\
 \phantom{F4}$r$ & 2.704 & 2.6623 & \phantom{+}$0.0691$ & $-0.0017$ &    ---    & &
 \phantom{F4}$i$ & 2.120 & 2.0927 & \phantom{+}$0.0464$ & $-0.0019$ &    ---    \\
 \phantom{F4}$z$ & 1.528 & 1.4713 & \phantom{+}$0.1576$ & $-0.1052$ &    ---    & &
             &       &        &              &           &           \\
\noalign{\vskip -0.1in} \\
\noalign{\centerline{\tt ubvri12}} \\
 \phantom{F4}$U$   & 4.801 & 4.3796 & \phantom{+}$1.3861$ & $-1.1400$ & $-0.0065$ & & 
 \phantom{F4}$B$   & 4.046 & 3.3504 & \phantom{+}$1.8900$ & $-1.2437$ & $-0.0143$ \\
 \phantom{F4}$V$   & 3.122 & 2.9364 & \phantom{+}$0.4604$ & $-0.2568$ & $-0.0022$ & &
 \phantom{F4}$R_C$ & 2.552 & 2.4248 & \phantom{+}$0.2672$ & $-0.0932$ &    ---    \\
 \phantom{F4}$I_C$ & 1.902 & 1.8662 & \phantom{+}$0.0657$ & $-0.0098$ &    ---    & &
               &       &        &              &           &           \\
\noalign{\vskip -0.1in} \\
\noalign{\centerline{\tt ubvri90}} \\
 \phantom{F4}$UX$  & 4.814 & 4.3241 & \phantom{+}$1.6005$ & $-1.3063$ & $-0.0073$ & &
 \phantom{F4}$BX$  & 4.032 & 3.2999 & \phantom{+}$2.0123$ & $-1.3425$ & $-0.0140$ \\
 \phantom{F4}$B$   & 4.049 & 3.3155 & \phantom{+}$2.0119$ & $-1.3400$ & $-0.0145$ & &
 \phantom{F4}$V$   & 3.129 & 2.9256 & \phantom{+}$0.5205$ & $-0.3078$ & $-0.0022$ \\
 \phantom{F4}$R_C$ & 2.558 & 2.4203 & \phantom{+}$0.3009$ & $-0.1220$ &    ---    & &
 \phantom{F4}$I_C$ & 1.885 & 1.8459 & \phantom{+}$0.0741$ & $-0.0151$ &    ---     \\
\hline
\end{tabular}
\begin{minipage}{1\textwidth}
$^{a}$~Based on the differences in the bolometric corrections for 
$E(B-V) = 0.0$ and $0.10$, assuming $\logg = 4.1$, 
$5250 \le \teff \le 7000$~K, $-2.0 \le \feh \le +0.25$, and 
$-0.4 \le \aFe \le +0.4$.  In the fitting equation for $R_\zeta$, 
$T_4 = 10^{-4}\,T_{\rm eff}$. \\
$^{b}$~Results are not provided for the $F218W$, $F225W$, $F275W$, and $F350LP$ 
filters because of their strong dependence on $T_{\rm eff}$.\\
\phantom{~~~~~~~~~~~~~~~~~~~~~~~}\\
\phantom{~~~~~~~~~~~~~~~~~~~~~~~}
\end{minipage}

\end{table*}

\section[]{Bias in a non-linear transformation}\label{bias}

Every time a non-linear transformation is applied to a variable that is subject 
to random errors (in this context e.g., an observed quantity with a Gaussian 
Probability Distribution Function; PDF) a skewed PDF arises as a consequence 
of the transformation. In this case the expectation value differs from the 
mode and the difference between the two introduces a bias. This traces 
directly back to a better known example, such as the bias between parallaxes 
and distances \citep[e.g.][]{lk73,c07}

Let us consider the case of transforming a magnitude $m$ into a heterochromatic 
flux $\mathcal{F}$, which (from Eq.~\ref{eq:pogson}) involves the following
non-linear relation
\begin{equation}\label{eq:mode}
\mathcal{F} = \alpha \,10^{-0.4\;m}
\end{equation}
where $\alpha$ is a constant that incorporates all of the details concerning 
the absolute calibration and zero-point of the photometric system under
consideration.

Assuming a Gaussian distribution for the errors around $m$ (as obtained e.g., 
by averaging different independent measurements)
\begin{equation}
f(\tilde{m}) = \frac{1}{\sqrt{2\pi}\sigma}\, 
               \rm{e}^{-\frac{(\tilde{m}-m)^2}{2\,\sigma^2}},
\end{equation}
the expectation value for the flux $\mathcal{F}$ is
\begin{displaymath}
E\left[\mathcal{F}|m\right] = \int_{-\infty}^{+\infty}\mathcal{F}(\tilde{m})\,
                              f(\tilde{m})\,\rm{d}\tilde{\it{m}}
\end{displaymath}
\begin{displaymath}
\phantom{E\left[\mathcal{F}|m\right]}=
\frac{\alpha}{\sqrt{2\pi}\sigma}10^{-0.4\,m}\int_{-\infty}^{+\infty}
10^{-0.4\,u}\rm{e}^{-\frac{{\it u}^2}{2\sigma^2}}\rm{d}{\it u}
\end{displaymath}
\begin{equation}\label{eq:expect}
\phantom{E\left[\mathcal{F}|m\right]} = \alpha\,10^{-0.4\,m}\,\rm{e}^{0.08\,\sigma^2\,\ln^2(10)}
\end{equation}
where $u=\tilde{m}-m$. The amplitude of the bias can be immediately estimated 
by comparing Eq.~(\ref{eq:expect}) and Eq.~(\ref{eq:mode}), or in fractional
terms
\begin{equation}
\frac{E\left[\mathcal{F}|m\right] - \mathcal{F}}{\mathcal{F}} = 
\rm{e}^{0.08\,\sigma^2\,\ln^2(10)} - 1.
\end{equation}
Fortunately, this difference is negligible in most investigations, when 
uncertainties are of the order of few hundredths of a magnitude, although it 
quickly rises from 1 percent at $\sigma \simeq 0.2$~mag to 10 percent at 
$\sigma \simeq 0.7$~mag.

Notice that the opposite bias is also present. In fact, in the above example 
we have assumed that the errors arise as a consequence of averaging a number of
independent measurements. In reality, when measuring magnitudes, several steps
are involved, which essentially consist of transforming number counts into
fluxes (for CCD detectors), 
and fluxes into magnitudes, which then are corrected for atmospheric extinction
(in case of ground-based observations) and standardized using photometric 
equations \citep[e.g.,][]{hfr81}. Assuming that the uncertainty of the 
photometric equations is small, which is usually the case if a sufficient number 
of standards has been observed spanning different colours and airmasses, with 
decreasing signal--to--noise the 
main uncertainty of a magnitude stems from its flux measurement. We also 
assume that 
the uncertainty in flux measurements is Gaussian, which is appropriate since a 
Poisson distribution is well approximated by a Gaussian after relatively few 
counts \citep[e.g.,][]{taylor_book}.

For objects with high signal--to--noise, the flux uncertainty has a narrow 
Gaussian peak, which transforms to negligible uncertainty in magnitude. 
However, as the signal--to--noise decreases, the uncertainty in the flux
broadens \citep[and other sources of random errors can also become 
important e.g.,][]{howell89}. As before, if we know the flux $\mathcal{F}$, 
we can compute the expectation value for a magnitude using the full PDF
\begin{displaymath}
E[m|\mathcal{F}]=\int_{-\infty}^{+\infty} m(\mathcal{\tilde{F}}) f(\mathcal{\tilde{F}}) \rm{d}\mathcal{\tilde{F}}
\end{displaymath}
\begin{displaymath}
\phantom{E[m|\mathcal{F}]} = -\frac{2.5}{\sqrt{2\pi}\sigma} \int_{-\infty}^{+\infty} \log\left({\frac{\mathcal{\tilde{F}}}{\alpha}}\right)\rm{e}^{-\frac{(\mathcal{\tilde{F}}-\mathcal{F})^2}{2\sigma^2}}\rm{d}\mathcal{\tilde{F}}
\end{displaymath}
\begin{equation}
\phantom{E[m|\mathcal{F}]} = 2.5 \log(\alpha) - \frac{2.5}{\sqrt{2\pi}\sigma} \int_{-\infty}^{+\infty} \log(\mathcal{\tilde{F}})\,\rm{e}^{-\frac{u^2}{2\sigma^2}}\rm{d}u
\end{equation}
where $u=\mathcal{\tilde{F}}-\mathcal{F}$, and expanding 
$\log(\mathcal{\tilde{F}})$ around the maximum of $-u^2$, we obtain 
\begin{displaymath}
E[m|\mathcal{F}] = 2.5 \log(\alpha) - \frac{2.5}{\sqrt{2\pi}\sigma} \int_{-\infty}^{+\infty} \Bigg[ \log(\mathcal{F}) 
\end{displaymath}
\begin{displaymath}
+ 5\log(\rm{e}) \sum_{n=1}^{\infty} \frac{(-1)^{n+1}}{n}\left( \frac{u}{\mathcal{F}} \right)^n \Bigg]\,\rm{e}^{-\frac{u^2}{2\sigma^2}}\rm{d}u
\end{displaymath}
\begin{equation}
=-2.5 \log\left(\frac{\mathcal{F}}{\alpha}\right) - 2.5\,\frac{5\log(\rm{e})}{\sqrt{2\pi}\sigma}\int_{-\infty}^{+\infty} \Bigg[ \sum_{n=1}^{\infty} \frac{(-1)^{n+1}}{n}\left( \frac{u}{\mathcal{F}} \right)^n \Bigg]\,\rm{e}^{-\frac{u^2}{2\sigma^2}}\rm{d}u
\end{equation}
whose solution is not trivial, especially when large errors will formally 
return negative fluxes. Nevertheless, for errors up to $\simeq30$\% in flux 
(i.e.~the limit where no negative flux is found within $3\sigma$), the above 
expression can be conveniently approximated using the first term of the 
expansion, thus leading to an estimate of the bias
\begin{equation}
E[m|\mathcal{F}] + 2.5 \log\left(\frac{\mathcal{F}}{\alpha}\right) \simeq 0.6 \left(\frac{\sigma}{\mathcal{F}}\right)^2.
\end{equation}

Notice that the two biases discussed here, $E[\mathcal{F}|m]$ and 
$E[m|\mathcal{F}]$ do not cancel each other, but are representative of two 
approaches typically used for estimating errors. In the first case, the bias 
in the flux arises if the error in magnitude is obtained by averaging different 
measurements. In the second, the bias follows from the fact that symmetric 
errors in the fluxes are asymmetric once transformed into magnitudes.
In either case, when publishing photometry, the procedure used to determine
the errors should always be described, in particular for the object with the
lowest precision (although formally the bias is always present). If this is 
not done, users should refrain from (or be very cautious in) analysing
statistically large samples of poor quality data, in the hope 
that average quantities (with formally small uncertainties) will still be 
useful: severe biases could undermine the mean results so obtained.

\end{document}